\documentclass[twocolumn]{aastex631}

\usepackage{amsmath,esint}
\usepackage{subfigure}
\usepackage{appendix}
\usepackage{bm}

\begin{document}

\title{Galactic Cosmic Ray Transport in the Giant Circumgalactic Medium Halo}

\author[0000-0002-0786-7307]{Chao-Ming Li}
\affiliation{Deutsches Elektronen Synchrotron (DESY), Platanenallee 6, D-15738 Zeuthen, Germany}
\affiliation{Humboldt University of Berlin, Unter den Linden 6, 10099 Berlin Germany}
\author[0000-0001-9473-4758]{Andrew M. Taylor}
\affiliation{Deutsches Elektronen Synchrotron (DESY), Platanenallee 6, D-15738 Zeuthen, Germany}
\correspondingauthor{Andrew M. Taylor}
\email{andrew.taylor@desy.de}


\begin{abstract}
Recent observations have revealed that the Milky Way is embedded in a massive circumgalactic medium (CGM) extending to several hundred kiloparsecs. Such an extended gaseous halo acts as both a reservoir of baryons and potentially as a confinement volume for Galactic cosmic rays (CRs). We investigate CR transport in this giant Galactic halo and compare its properties with those of conventional small-halo models. In the giant-halo scenario, the halo height is no longer a free parameter, but instead relates to the extent of the source region. We show that CR transport within the source region remains similar to that in small-halo models, while substantial differences emerge at larger distances. In the giant-halo scenario, CRs develop an extended approximately 1/r spatial tail and exhibit a broader age distribution than the small-halo case. This model is shown to be consistent with current secondary-to-primary CR measurements. We further find that uncertainties associated with Galactic gas distributions are comparable to those arising from nuclear spallation cross sections. These results suggest that the giant-halo model provides a physically motivated alternative to conventional small-halo models and may have important implications for diffuse gamma-ray and neutrino emission from the Galactic environment.

\end{abstract}



\section{Introduction} \label{sec:intro}
Cosmic rays (CRs) are particles accelerated up to (ultra-)relativistic energies that propagate throughout the Universe. Owing to their electric charge, CRs are readily deflected by magnetic fields, resulting in a near–random-walk propagation that is commonly described as diffusion. During their transport from their sources to the Earth, primary CRs can interact with the intervening gas, producing secondary CRs through spallation processes. The spectral energy distributions (SEDs) of primary and secondary CRs, as well as their ratios, are key observables for probing CR propagation mechanisms. In particular, secondary-to-primary ratios are governed by the integrated column density of target material encountered along CR trajectories, known as the grammage, defined as $\chi = \int n_H dl$, where $n_H$ is the gas density along the CR trajectory. In most propagation models, the interstellar medium (ISM) is treated as the primary target for CR interactions, as implemented in widely used open-source codes such as GALPROP\footnote{\url{https://galprop.stanford.edu/}} and DRAGON2 \citep{Dragon_I, dragon_II_2018}. 

The ISM is mainly composed of molecular clouds ($\rm H_2$), atomic gas ($\rm HI$) and ionized gas ($\rm HII$), with a total mass of $\rm \sim 10^{10} \, M_{\odot}$ \citep{Cautun_2020}. Its mean number density can vary from $\rm 0.1-10 \, cm^{-3}$, implying a characteristic diffusion length of $\rm \lesssim 1 \, kpc$ for CRs before undergoing spallation\footnote{The distribution of molecular clouds is highly clumpy, with local densities reaching $\rm 10 \sim 1000 \, cm^{-3}$, leading to significantly different local environments. Nevertheless, CR propagation simulations usually adopt a smoothed, averaged gas distribution for computational simplicity. }. The spatial distribution of the ISM largely follows the Galactic disk, with scale heights in the vertical (Galactic latitude) direction ranging from tens to hundreds of parsecs (pc) \citep{Cox_2005}. These characteristics motivate the widely used leaky-box models and diffusive halo with thin disk gas models for Galactic CR propagation \citep{Strong_2007, Blasi_2012, Evoli_2019,Genolini_2019, Evoli_2020}. 
In these models, CRs diffuse within a finite confinement volume, interact with gas within a subset of this volume, and are assumed to escape permanently once they reach an absorbing boundary located several kiloparsecs above the disk, which is referred to as the halo height. Despite their simplicity, these models are able to reproduce some key observational results, including those from AMS-02 \citep{AMS_2021} to a reasonable extent.

However, such an absorbing boundary is artificially introduced and has no clear  physical origin. Furthermore, as \cite{Hopkins_2022} has argued, the commonly inferred “halo height” may instead be interpreted as the region within which CRs have an order-unity probability of returning to the Solar neighborhood, rather than as a sharp physical boundary separating confined and unconfined particles. This raises the question of whether Galactic CR transport should be considered within a substantially larger volume than is usually assumed.

At the same time, observations over the last decade have revealed that  the circumgalactic medium (CGM) is a massive reservoir of baryons and accounts for a large fraction of the missing baryon problem \citep{ Weiner_1996, Tumlinson_2017,Bregman_2022,Zhang_2024}. The baryons in the CGM have a multiphase distribution in their ionization structure, with the hot component containing $\rm \sim 1\times10^{11} \, M_{\odot}$ \citep{Zhang_2024,Zhang_2026, Bregman_2022}, which exceeds the stellar mass of the Milky Way \citep{Cautun_2020}. Consequently, the CGM hosts a substantially larger number of hydrogen nuclei than the ISM, providing an abundant reservoir of potential interaction targets for CRs. 

The CGM may also play an important role in the long-term evolution of Galactic CRs. The characteristic residence time around the galactic disk inferred from CR composition measurements is of the order $10^7$ years, whereas the Milky Way has likely continuously injected CRs for at least $10^{9}$ years. If CRs are able to diffuse throughout the extended Galactic halo, a large population of particles should accumulate in the CGM over cosmic time. Although the magnetic field structure of the CGM remains poorly constrained \citep{Mora_2019, Heesen_2023, Ramesh_2023}, the low gas density ($n_{\rm H}\sim10^{-4} \,\mathrm{cm^{-3}}$) implies long spallation timescales and allows CRs to propagate to distances far beyond the conventional diffusive halo region.

Motivated by these developments, in this paper we investigate CR transport in a giant Galactic halo extending out to the virial radius of the Milky Way. This paper is organized as follows. In Sec.~\ref{sec:model} we describe the transport model and numerical methods. In Sec.~\ref{sec:results} we present the resulting CR spatial distributions, age distributions, and composition ratios. In Sec.~\ref{sec: discussion} we discuss the implications of the giant-halo scenario and its observational consequences. Our conclusions are summarized in Sec.~\ref{sec:conclusion}.

\section{Model description}
\label{sec:model}

\begin{figure*}[htbp]
\centering
\includegraphics[width=0.8\textwidth]{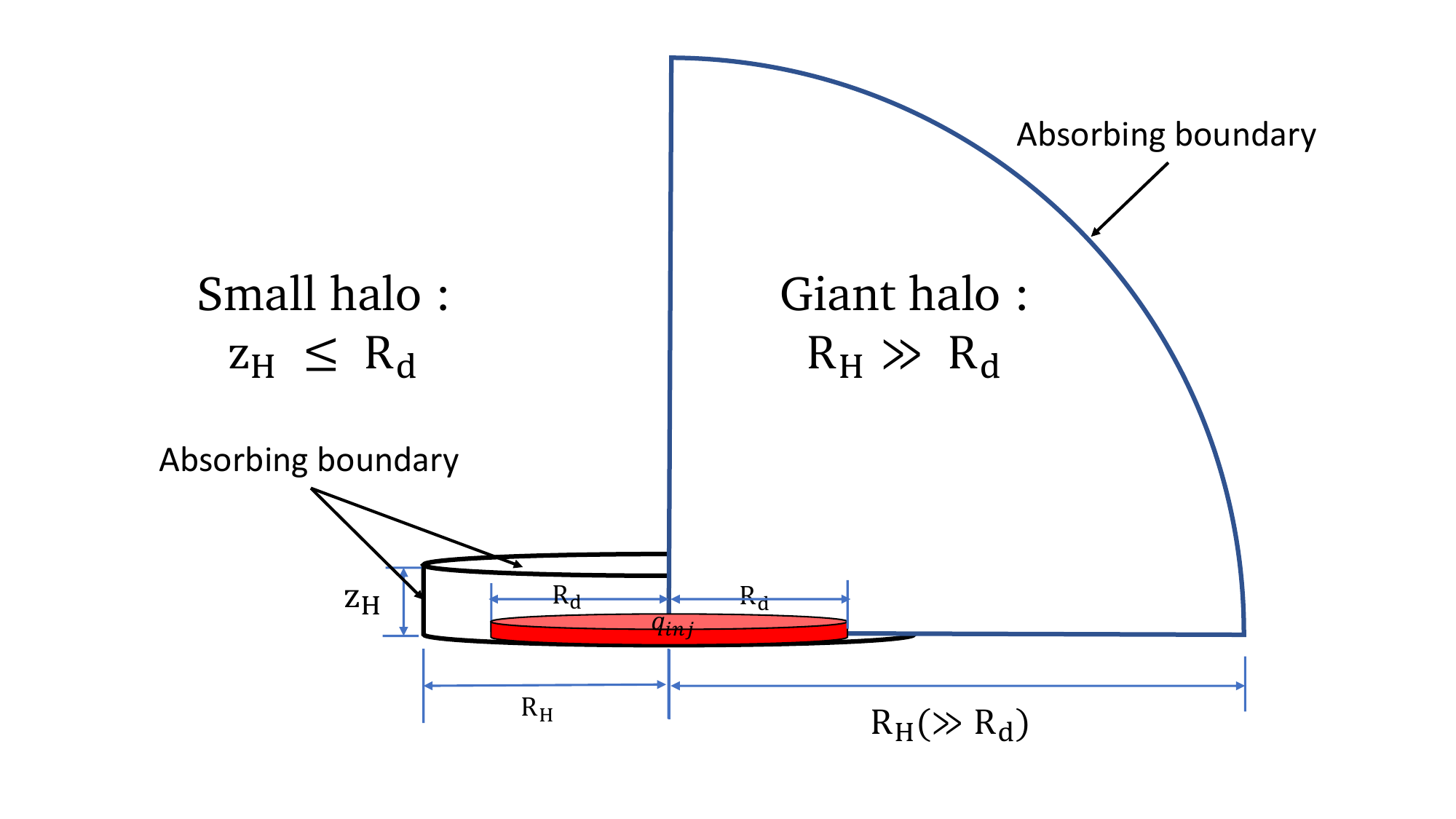}
\caption{The comparison of the small-halo scenario and the giant-halo scenario. The red thin disk represents the CR injection source and $\rm R_d$ is the source radius. $\rm R_H,\, z_H$  are the radius of absorbing boundaries in $\rm R$ and $\rm z$ direction.}
\label{fig:model}
\end{figure*}

\subsection{Cosmic Ray Transport Equation}\label{subsec: equation}

In this work we consider the rigidity range 10–100 GV, where CR transport is expected to be dominated by spatial diffusion. Energy losses, diffusive reacceleration, and solar modulation have negligible effects and are therefore neglected. We further assume that large-scale Galactic winds do not dominate transport within the region of interest. Under these assumptions the CR transport equation reduces to:
\begin{equation}\label{eq:Diffusion_equation_0}
\begin{aligned}
    \frac{\partial n_i}{ \partial t}  &=  D\nabla^2 n_i +  \Sigma_j \Big (\frac{n_j}{\tau_{s,j\to i}}+ \frac{n_j}{\gamma\tau_{d,j}} \Big ) \\
    &- \frac{n_i}{\tau_{s,i}} -\frac{n_i}{\gamma\tau_{d,i}} + q_i,
\end{aligned}
\end{equation}
where $n_i = dN / d^3r $ is the density of CRs at rigidity ($\mathcal{R} = pc/Ze$), $n_j$ denotes a heavier parent species that can produce $n_i$ through spallation or decay, and $\tau_{d,j}$ is its decay timescale ($\tau_d=\infty$ for stable species). The spallation timescale of species $i$ is $\tau_{s,i}=(n_H\beta c\sigma_i)^{-1}$, while the production timescale for the channel $j\to i$ is $\tau_{s,j\to i} = 1 / (n_H \beta c\sigma_{j\to i})$, where $\sigma_{j\to i}$ is the corresponding spallation cross section. The target gas density is denoted by $n_H$\footnote{Hydrogen constitutes at least 90 \% of ISM, while 10 \% of ISM is made up of Helium and heavier elements which also provide targets for CR spallation. Their contribution can be effectively included through an enhancement factor \citep{Mori_2009}  }. The source injection rate is $q_i$, and it is assumed to be time independent. For simplicity, we adopt a spatially uniform diffusion coefficient throughout both the Galactic disk and the CGM: $D(\mathcal{R})=\beta D_0(\mathcal{R} / GV)^\delta$, with $\beta$ the CR velocity in units of the speed of light, $D_0$ the diffusion coefficient at 1 GV, and $\delta$ the energy dependence of diffusion coefficient. 

\subsection{Boundary conditions}

Fig. \ref{fig:model} compares the setups of the small- and giant-halo scenarios. In both cases, the CR source is modeled as a uniform disk. In the classical small-halo case, absorbing boundaries ($n_{\rm CR}=0$) are placed close to the source, making CR diffusion effectively one-dimensional in the $z$ direction; analytical solutions have been derived in previous studies \citep{Morrison_1954,Lerche_1982,Putze_2010}. In contrast, the giant-halo case places the absorbing boundary near the virial radius, much larger than the source size, so CR transport is fully three-dimensional. The corresponding simplified 3D diffusion solution is given in Appendix \ref{sec: appendix_1}.

\subsection{Differential equation solver}\label{subsec: numerical method}

To solve Eq.~(\ref{eq:Diffusion_equation_0}) in realistic Galactic environments, we employ both finite difference and Monte Carlo methods.

\subsubsection{Finite difference method}
We solve the spallation network including $\rm {}^{16}O$, $\rm {}^{15}N$, $\rm  {}^{14}N$, $\rm {}^{13}C$, $\rm {}^{12}C$, $\rm {}^{11}B$, $\rm {}^{10}B$, $\rm {}^{10}Be$, $\rm {}^{9}Be$ and $\rm {}^{7}Be$\footnote{The half-life of $\rm {}^{7}Be$ in laboratory conditions is only tens of days, and it decays via electron capture from its inner shells. For relativistic $\rm {}^{7}Be$ in cosmic rays, all electrons have been stripped, so they can be treated as stable secondary CRs \citep{Yiou_1970, Borchiellini_2026}.}. Among them, $\rm {}^{16}O, {}^{14}N, {}^{12}C$ are treated as primary CRs injected from sources with relative abundances of $1:0.1:1$ \citep{AMS_2021}. All the other species are treated as secondary CRs, produced either directly or indirectly from the injected primaries. Although heavier CRs like $\rm Ne, Mg,$ and $\rm Si $ can also contribute to these CRs, the network considered here accounts for approximately $80\% -90\%$ of their total production \citep{dragon_II_2018, Genolini_2015, Mewaldt_1981}, providing a reasonable balance between completeness and computational efficiency. 

The Crank–Nicolson method, combined with the Alternating Direction Implicit (ADI) scheme, is employed to solve Eq. (\ref{eq:Diffusion_equation_0}) in a cylindrical coordinate system, and azimuthal symmetry is assumed for simplicity. Since the gas density varies by several orders of magnitude: from $\rm \sim 1 \, cm^{-3}$ on the disk to $\rm \sim 10^{-4} \, cm^{-3}$ close to the virial radius, we adopt a logarithmic spatial grid and ensure that $\rm dx < \min(z_{gas}, \sqrt{D\tau})$ where $\rm z_{gas}$ is the scale height of gas distribution and $\rm \sqrt{D\tau}$ is the characteristic diffusion length before decay or spallation. CR sources are distributed uniformly in a thin Galactic disk.

For the boundary condition, we set an absorbing boundary at $\rm R_H = 20 \, kpc,\, z_H \le R_d$ for the small-halo model, and  $\rm R_H = z_H \gg R_d$ for giant-halo model. Convergence was verified by doubling the spatial and (or) time  resolution and confirming that the resultant CR fluxes changed by less than 1\% in all cases.

\subsubsection{Monte Carlo simulation}

Monte Carlo simulations can provide complementary information to that of the finite difference method, such as the age distribution of CRs reaching Earth. We inject 5000 particles uniformly from a infinitely thin Galactic disk and follow their three-dimensional random walks. A particle is considered detected when it enters a spherical region of radius $\rm R = l_{scatter}$ centered on Earth. Assuming azimuthal symmetry, the detection region can be replaced by a torus of minor radius $\rm l_{scatter}$ centered at Earth's Galactocentric radius, significantly increasing the number of detected particles and thus improving the detection statistics. Each simulation is run for a fixed duration (see Fig. \ref{fig: age}). To model continuous CR injection, detected particles are recorded at every timestep throughout the simulation. The age distribution is then constructed by summing the detected particles over all timesteps, which is equivalent to superposing CR contributions from sources continuously injecting over time. Random numbers are generated using the thread-safe Mersenne Twister Dynamic Creator\footnote{\url{https://www.math.sci.hiroshima-u.ac.jp/m-mat/MT/emt.html}}. 

For the boundary condition, we do not set a boundary for the giant-halo scenario, because CRs below 100~GV can hardly escape from the Milky Way within the Hubble time. Instead, a termination timescale for the simulation in this case is adopted. For the small-halo scenario, we set the absorbing boundary at $\rm R_H = 20 \, kpc, z_H = 10 \, kpc$. Once a particle has escaped from this region, we remove it from the simulation. 

\begin{figure*}[htbp]
\centering
\includegraphics[width=0.8\textwidth]{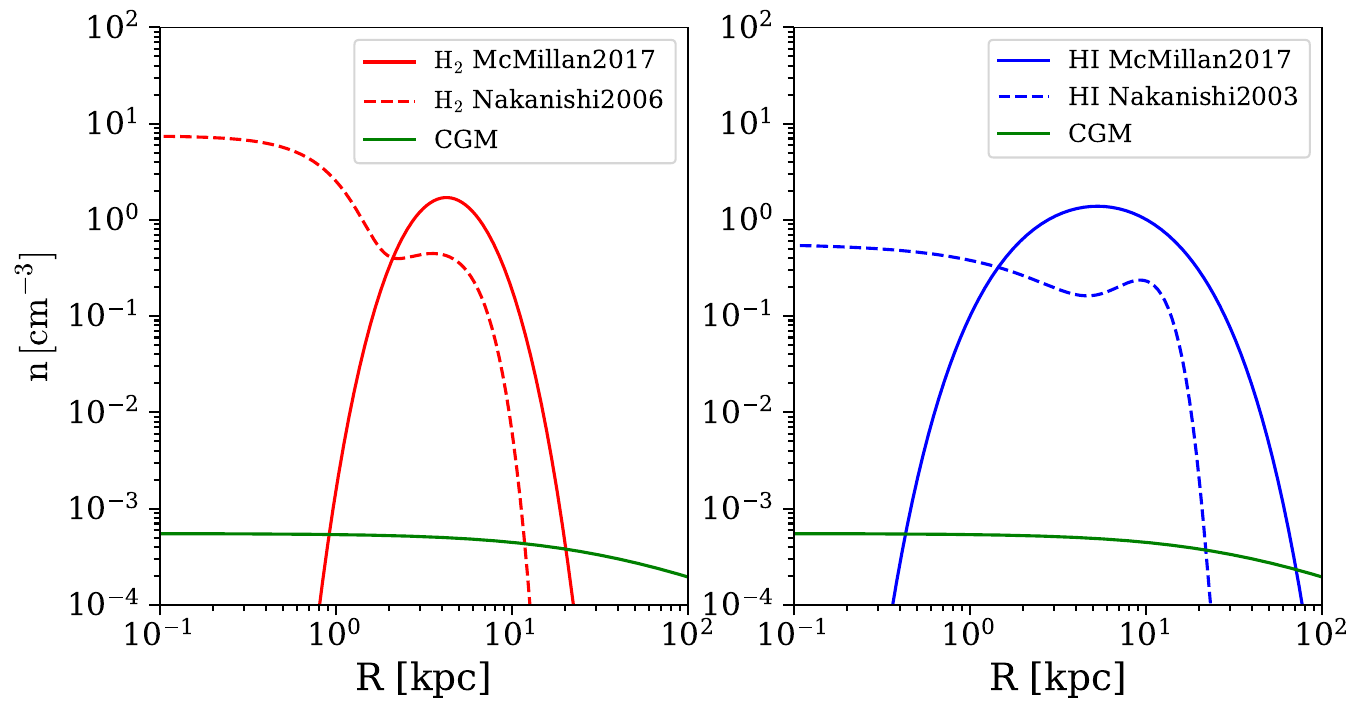}
\caption{Comparison of gas density models as a function of galactocentric radius in the Galactic plane. The left panel shows molecular gas, while the right panel shows atomic hydrogen. The CGM distribution is shown in both panels for comparison.}
\label{fig:gas}
\end{figure*}

\subsubsection{Gas distribution}\label{subsec: gas}

We include three gas components for the Milky Way gas model: $\rm H_2$, HI, and CGM. The disk is dominated by $\rm H_2$ and HI, while the CGM becomes important at large radii ($\gtrsim20\,{\rm kpc}$).

For the ISM gas model, we adopt the analytical model from \cite{McMillan_2017}, based on dynamical constraints from Galactic observations. In this model, the total masses of $\rm H_2$ and HI are approximately $\rm 10^9\,M_\odot$ and $\rm 10^{10}\,M_\odot$, respectively. The central molecular zone, with mass $\sim6\times10^7\,M_\odot$ \citep{Ferriere_2007}, is neglected because its contribution is small compared to the disk gas.

Compared with the Nakanishi2003/2006 model which is used in other CR simulations \citep{Dragon_I}, the McMillan2017 model predicts a denser and more extended ISM gas distribution (Fig. \ref{fig:gas}). One possible reason for this difference is that the Nakanishi2003/2006 model is derived primarily from HI/CO surveys, which may not trace the full ISM content. In order to evaluate the impact of gas uncertainties on the CR propagation results, we consider both of these gas models.

Observations of Galactic rotation curves \citep{Klypin_2002,Bhattacharjee_2014} suggest an approximately logarithmic halo potential, which in hydrostatic models implies a power-law CGM density profile \citep{Faerman_2017, Tepper_2015}. Similar profiles are also predicted by galaxy simulations \citep{Maller_2004}. A recent notable result is from \cite{Zhang_2024}, who analyzed the thermal X-ray emission from CGM with extended ROentgen survey with an Imaging Telescope Array (eROSITA) and fit the density profile of hot gas with a $\beta$ function. However, this model does not account for the multiphase CGM baryons. We therefore adopt the hydrostatic CGM model of \cite{Maller_2004}, which predicts a total halo gas mass of $\sim10^{11}\,M_\odot$ and a somewhat shallower radial decline.

\subsubsection{Spallation cross sections}\label{subsec: cross section}

\begin{figure}[htbp]
\centering
\includegraphics[width=0.4\textwidth]{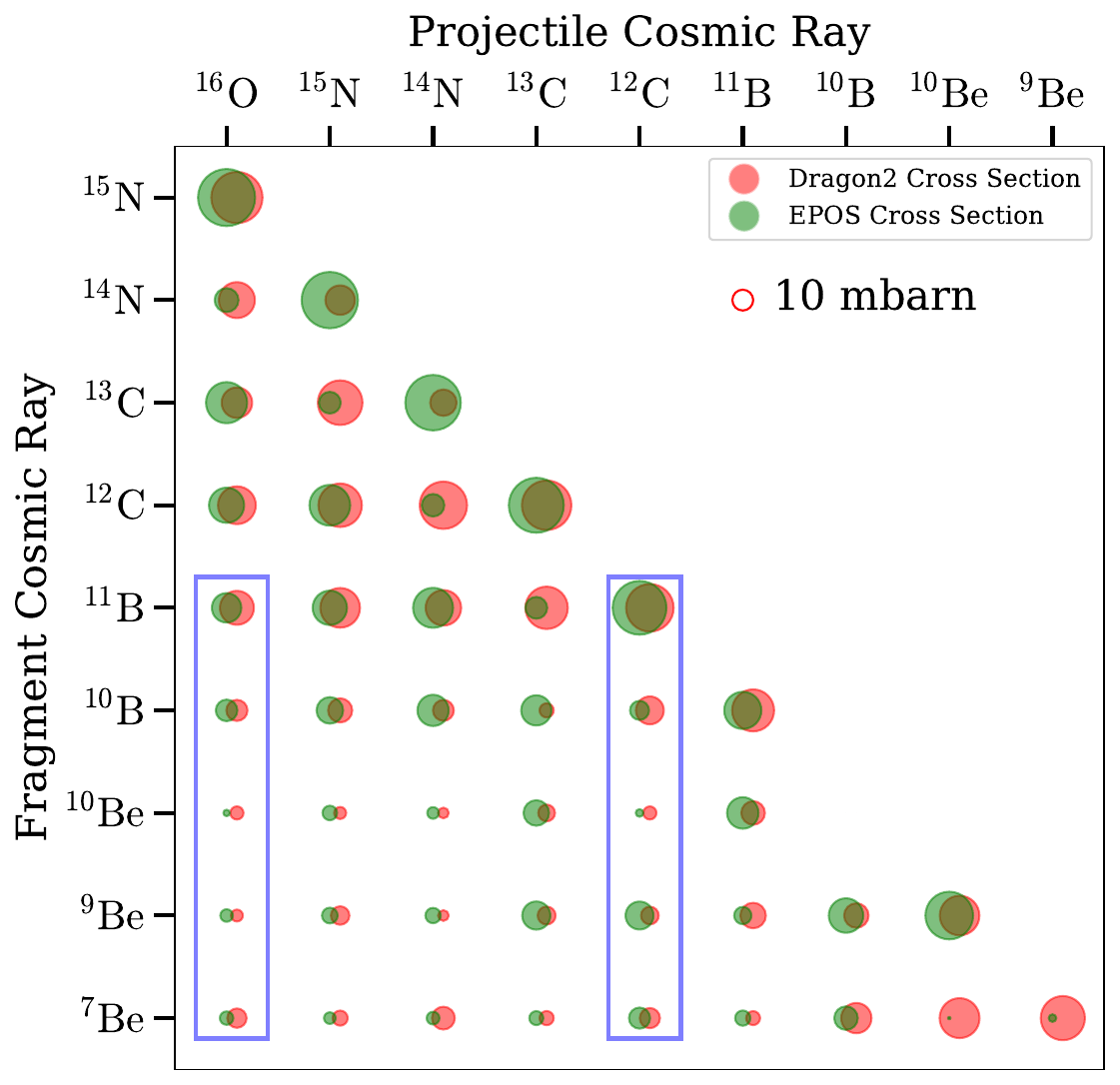}
\caption{The comparison of the spallation cross sections from DRAGON2 and EPOS model in CRMC software. The area of the circles represents the absolute value of the cross sections, and the reference value of 10 mbarn is shown below the legend. Channels dominating the production of boron and beryllium isotopes are highlighted by blue rectangles. }
\label{fig:cross section}
\end{figure}

Nuclear interaction cross sections play a key role in the production of secondary CRs, and their uncertainties remain one of the dominant sources of uncertainty in CR propagation models \citep{Tomassetti_2017,Strong_2007,Evoli_2018}. We use the Cosmic Ray Monte Carlo (CRMC) framework to simulate the differential cross sections for interactions between CR nuclei and target hydrogen nuclei \citep{Ulrich_2021}. Since the inelastic cross sections increase logarithmically with energy \citep{Fagundes_2012}, simulations are performed in multiple energy bins, with $10^5$ events generated in each energy bin. In these simulations, CR isotopes are treated as projectiles and fired at a fixed target composed of protons. The production cross section of a given isotope is computed as $\sigma = \sigma_{tot}\times N_{isotope} / N_{tot} $, where $\sigma_{tot}$ is the total inelastic cross section, $N_{isotope}$ is the number of events producing the isotope of interest and $N_{tot}$ is the total number of simulation events. Note that the values of $N_{isotope}$ here correspond to the "effective" production cross sections \citep{Tomassetti_2017}, which account for contributions from short-lived intermediate nuclei.

Although several hadronic interaction models implemented in CRMC—such as EPOS-LHC, QGSJET, and DPMJET—provide consistent total inelastic nuclear cross sections, the isotope-specific production cross sections for particular channels (i.e. $\rm {}^{12}C + p \to {}^{11}B$) still vary from model to model. In this work, we adopt the EPOS-LHC model in CRMC and compare the production cross sections with those in DRAGON2 code (Figure~\ref{fig:cross section}). The total production cross sections of boron ($\rm {}^{11}B + {}^{10}B$) and beryllium ($\rm {}^{10}Be + {}^{9}Be + {}^{7}Be$) are dominated by channels from primary $\rm {}^{16}O$ and $\rm {}^{12}C$, for which the two models give similar results. However, for the production of radioactive $\rm {}^{10}Be$, DRAGON2 predicts that primary species produce about three times more than EPOS-LHC, while secondary species produce significantly less. As a result, it is unclear whether the two models yield similar overall $\rm {}^{10}Be$ production. We will discuss $\rm {}^{10}Be$ production in detail in the discussion section (see \ref{sec: discussion}).

\section{Results}
\label{sec:results}

\subsection{Small halo VS giant halo}

In conventional simulations of CR transport in the diffusive halo, the halo size is typically limited to  $\rm R_H\lesssim 20 \, kpc, \,z_H\lesssim 10 \, kpc$. However, to investigate CR propagation in the CGM, the halo must be extended out to scales comparable to the virial radius of the galaxy. We therefore adopt a much larger computational domain out to the Galactic virial radius for our giant-halo model. 

\subsubsection{Cosmic ray spatial distribution}
\begin{figure*}[htbp]
\centering
\includegraphics[width=1.0\textwidth]{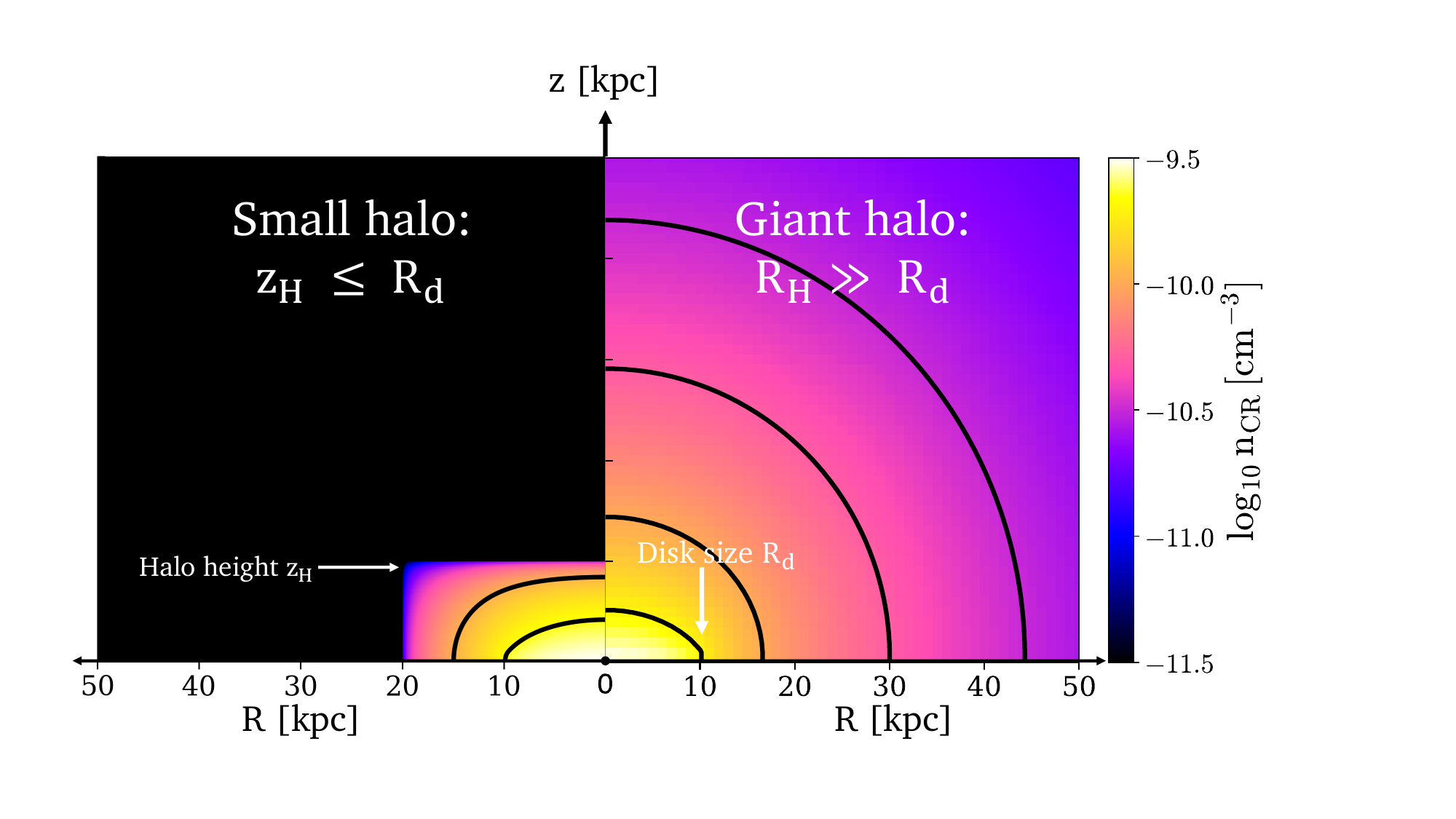}
\caption{Comparison of primary CR density distribution for small halo (left) and giant halo (right). The free escape boundary for small halo is $\rm z_H = 10 \, kpc, \, R_H = 20 \, kpc$). For the giant halo, the free escape boundary is ($\rm R_H =  250 \, kpc$). Here we only plot the inner region up to 50 kpc to compare with the small halo. The colorbar is re-normalized to be close (not fit) to the typical CR density. }
\label{fig:Compare_1D_3D}
\end{figure*}

\begin{figure}[htbp]
\centering
\includegraphics[width=0.4\textwidth]{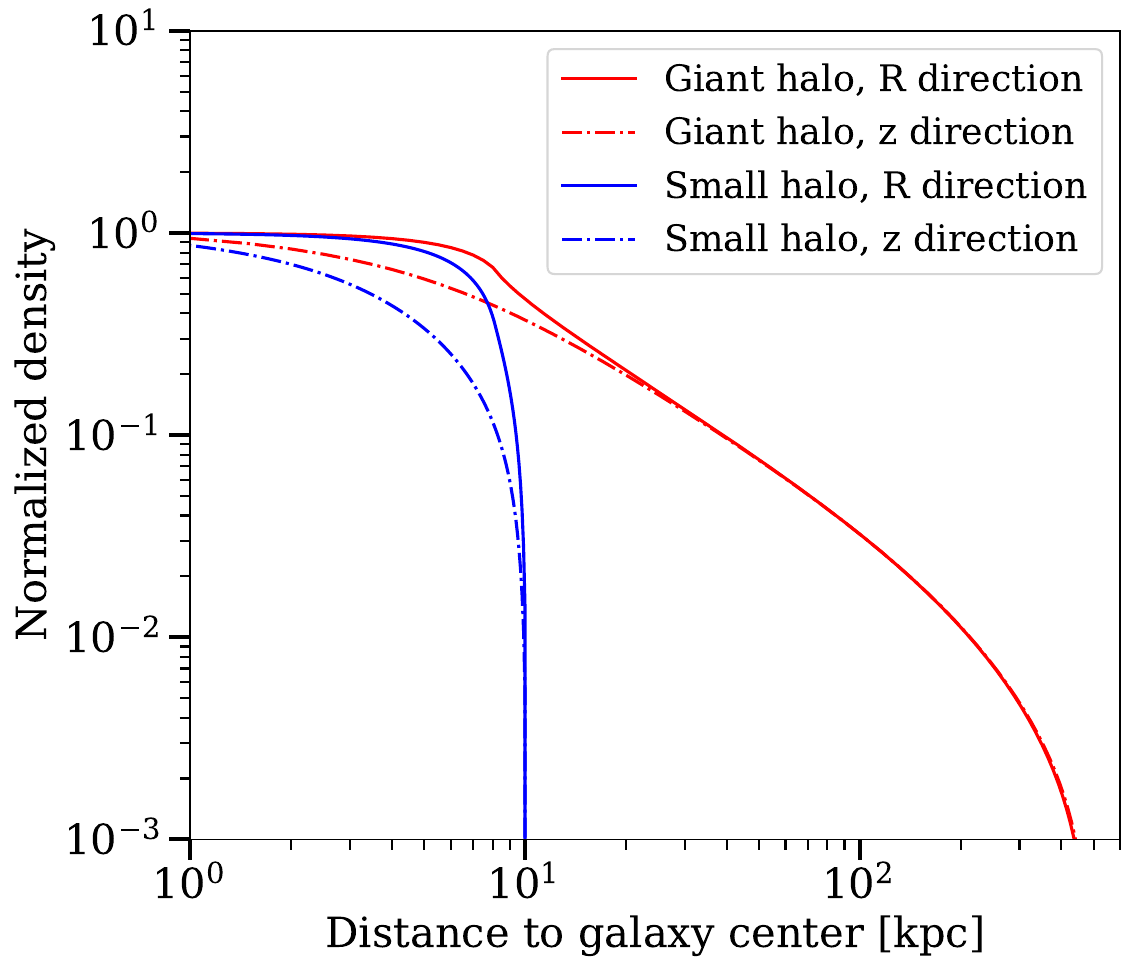}
\caption{Cosmic ray distribution due to diffusion for small-halo and giant-halo scenario. The parameter settings are:  $\rm R_d = 10 \, kpc$ for both cases;  $\rm R_H = 10 \, kpc \, ,z_H = 10 \, kpc$ for small-halo scenario and $\rm R_H = 250 \, kpc$ for giant-halo scenario. } 
\label{fig: radial profile}
\end{figure}

Fig. \ref{fig:Compare_1D_3D} compares the steady state CR density distribution obtained for both the small- and giant-halo scenarios, assuming diffusion-only transport.

In the small-halo case, CRs remain concentrated around the Galactic disk and interact predominantly with the ISM. The imposed absorbing boundaries lead to a sharp truncation of the CR density at the halo height, because particles crossing an absorbing boundary escape permanently from the system. This boundary effect leads to a spherically asymmetric CR distribution, and is necessarily imposed in order that steady state density distribution be achievable for the 1D-like diffusion.

In the giant-halo scenario the CR distribution is allowed to extend out to the Galactic virial radius. While, as in the small-halo case, it remains spherically asymmetric at small distances — reflecting the underlying source morphology — the CRs gradually "forget" their source details further out. At larger distances, their distribution asymptotes towards a spherically symmetric distribution. In this case, steady state is achieved initially at smaller radii, with the radius out to which steady state holds increasing with time. For the large halo, the steady state density distribution is achieved due to the growth in phase space with distance (only possible for the 3D case) rather than due to the imposition of the absorbing boundary.

Fig.~\ref{fig: radial profile} compares the normalized CR profiles in the radial ($R$) and vertical ($z$) directions for the two scenarios. To facilitate comparison, the halo height in the small-halo model is chosen to match the source size in the giant-halo model. In the small-halo case, the CR density is nearly uniform before sharply dropping at the absorbing boundary. For the giant-halo scenario the CR distribution is flat at radii smaller than the source region, resembling the small-halo scenario in this region. However at radii larger than the source region, the profile has a very long $1/r$ tail, characteristic of 3D diffusion at steady state. 

The spallation attenuation length in the CGM can be estimated as $\rm \sqrt{D\tau} \sim 67 \, kpc \, (D / 3\times10^{28} cm^2 s^{-1}) (n_H / 10^{-4} cm^{-3})^{-1} (\sigma / 237 \, mbarn)^{-1}$, indicating that the $1/r$ CR density profile can persist out to tens or even hundreds of kpcs, provided that the diffusion timescale does not exceed the Hubble time. 
For such a profile, the differential number of CRs contained within a spherical shell, $dN_r = 4\pi r^2 n_{CR} dr$ increases linearly with radius, implying that the majority of CRs actually reside out at large distances from the Galactic disk at late times.

\subsubsection{Cosmic ray age distribution}
\begin{figure*}[htbp]
\centering
\includegraphics[width=1.0\textwidth]{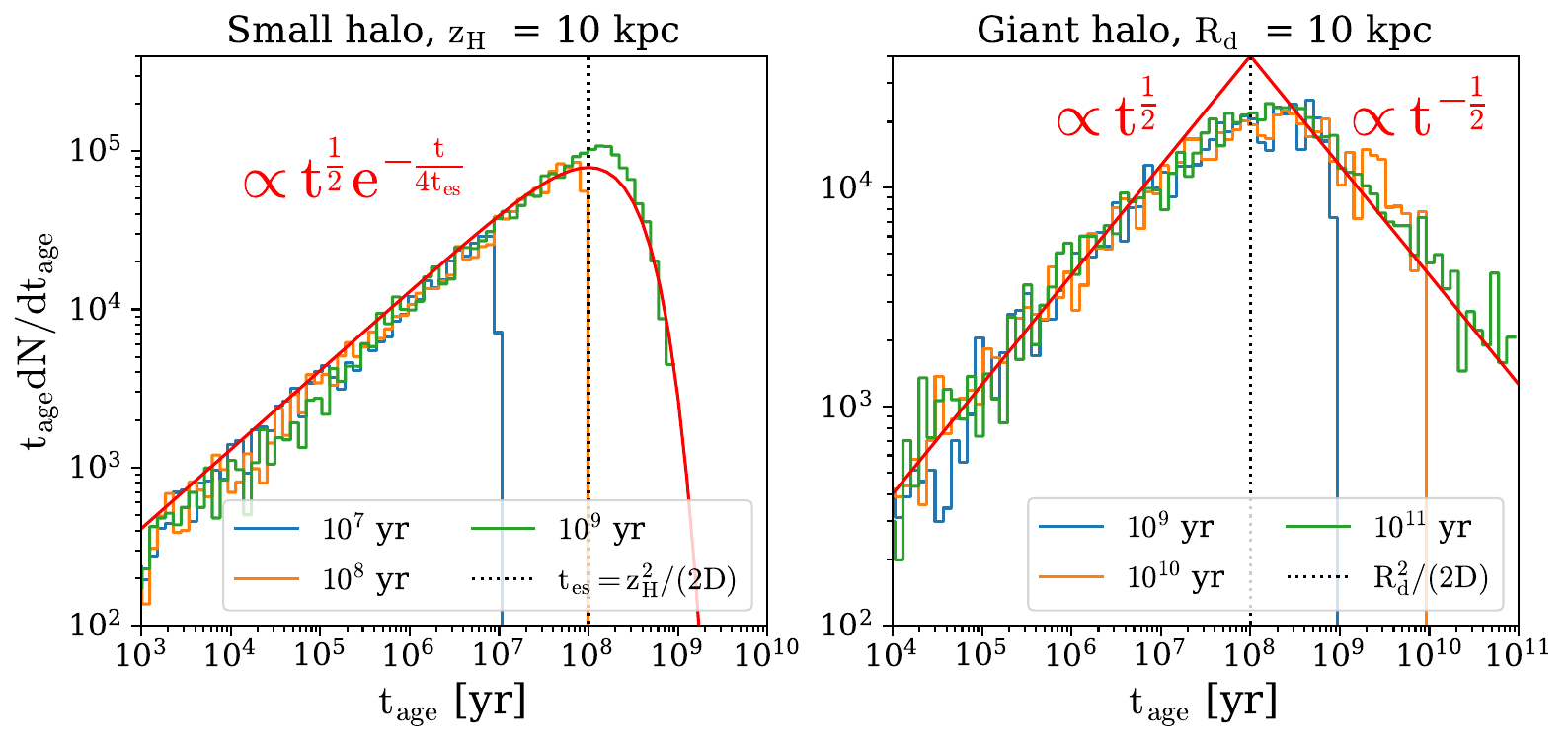}
\caption{Particle age distribution at Earth from Monte Carlo simulation for both small halo (left panel) and giant halo (right panel) cases. Different colors of histograms represent different simulation time. The red curves show the fitting with analytical formulas.  }
\label{fig: age}
\end{figure*}

The age distribution of CRs observed at Earth ($\vec{r}_E$) describes the distribution of propagation times in the Galaxy, a quantity particularly relevant for radioactive species with finite lifetimes. It can be derived from the Green's function for the case of diffusive transport from a point source:
\begin{equation}
\mathcal{P}(\tau|\vec{r_E}) = \frac{G(\vec{r_E},\tau)}{\int_0^\infty G(\vec{r_E},\tau')d\tau'}
\end{equation}
where
\begin{equation}
G(r,\tau) = \frac{1}{(4\pi D\tau)^{d/2}} \exp(-\frac{r^2}{4D\tau})
\end{equation}
with $d$ the dimensionality of the diffusion setup and $D$ the diffusion coefficient. The resulting age distribution is
\begin{equation}\label{eq:age distribution}
    \mathcal{P}(\tau|\vec{r_E}) \propto \left \{
\begin{aligned}
    & \tau^{-1/2} \exp(-\frac{\vec{r_E}^2}{4D\tau}), \quad d = 1 \\
    & \tau^{-3/2} \exp(-\frac{\vec{r_E}^2}{4D\tau}), \quad d = 3
\end{aligned}
\right.
\end{equation}
The exponential suppression at small $\tau$ reflects the finite travel time required for CRs to reach Earth, while at large $\tau$ the distribution approaches a power law with index $-d/2$. For spatially extended source distributions, integrating Eq.~(\ref{eq:age distribution}) over space preserves the power-law time dependence at large $\tau$. In the small-halo case, however, the absorbing boundary introduces an additional exponential cutoff at large ages \citep{Lipari_2014, Lipari_2022}.

Fig. \ref{fig: age} shows the results from our Monte Carlo simulation for both small-halo and giant-halo scenarios. In the small-halo case, the age distribution follows $dn/dt \propto t^{-1/2}\exp(-t / 4t_{es})$, consistent with 1D diffusion with absorbing boundary. The location of the cutoff is determined by the diffusive escape timescale. Note that our analytical fit expression in this figure (see red line) shares the same form but different parameters with \cite{Lipari_2022}, reflecting that our small-halo geometry approximates to — but does not exactly reproduce — the 1D diffusion model result.

In the giant-halo case, the age distribution is a broken power law, with the break timescale set by the source size. At early times, CR propagation is effectively 1D, giving: $dn/dt \propto t^{-1/2}$. At later times, the propagation becomes fully 3D and the distribution steepens to : $dn/dt \propto t^{-3/2}$. This transition is a key feature of the giant-halo model, which predicts a larger population of old CRs at Earth compared to the small halo model, especially important for the radioactive CRs such as $\rm ^{10}Be$.

The giant-halo picture is potentially supported by \cite{Lipari_2022}, who found that preliminary AMS-02 $\rm {}^{10}Be / {}^{9}Be $ data favor a broader CR age distribution and a more extended galactic confinement volume. 

\subsubsection{Halo height and disk size}

As discussed above, the source size acts as a critical scale in the giant-halo scenario. As demonstrated in Fig.~\ref{fig: age}, before CRs diffuse beyond the source region, propagation is effectively describable by 1D transport, with the particles having an $\mathcal{O}(1)$ return probability for returning back to the source region.  
Beyond the source region scale, however, diffusion becomes fully 3D, resulting in the particle return probability to the source region dropping significantly below $\mathcal{O}(1)$.  

The diffusion history in the giant-halo scenario also reflects this transition between 1D and 3D like propagation. Following their continuous injection in the source region, CRs initially accumulate and reach steady state within the source region on a timescale $t \sim R_d^2 / D$, ie. the source region size acts in a similar manner to the halo size in the small-halo case. During this stage, both the spatial and age distributions for the giant halo case are similar to that for the small-halo scenario. On longer timescales than this, CRs start to "leak" to the outer giant-halo region, where their propagation becomes genuinely 3D, and their return probability drops significantly below $\mathcal{O}(1)$. 

One advantage of giant-halo scenario is that it reproduces the small-halo behavior within the source region and naturally incorporates CR interactions with the CGM. These interactions with the CGM are particularly relevant for accounting for the grammage which CR, arriving to Earth, pick up. In addition, the halo height is no longer treated as a free parameter, and the source size characterises the transition distance at which CRs diffusion shifts from effectively 1D to fully 3D. The source size can potentially be constrained observationally by the CR source candidate distribution, such as supernova remnants \citep{Blasi_2013} and micro quasars \citep{LHAASO_2025}.

\subsection{Secondary to primary ratio}

The secondary to primary CR ratios carry information of CR transport in the galaxy. Stable secondary to primary CR ratios (i.e. B/C) are directly related to the grammage that CR have picked up during their propagation. Unstable secondary to primary CR ratios (i.e. Be/C) also provide complimentary information about the CR confinement time within the galaxy. Here we determine the B/C and Be/C ratios, for both small-halo and giant-halo models, and show how the uncertainties in cross sections and gas templates affect the results.

\subsubsection{Small-halo scenario}
\begin{figure*}[htbp]
\centering
\includegraphics[width=0.8\textwidth]{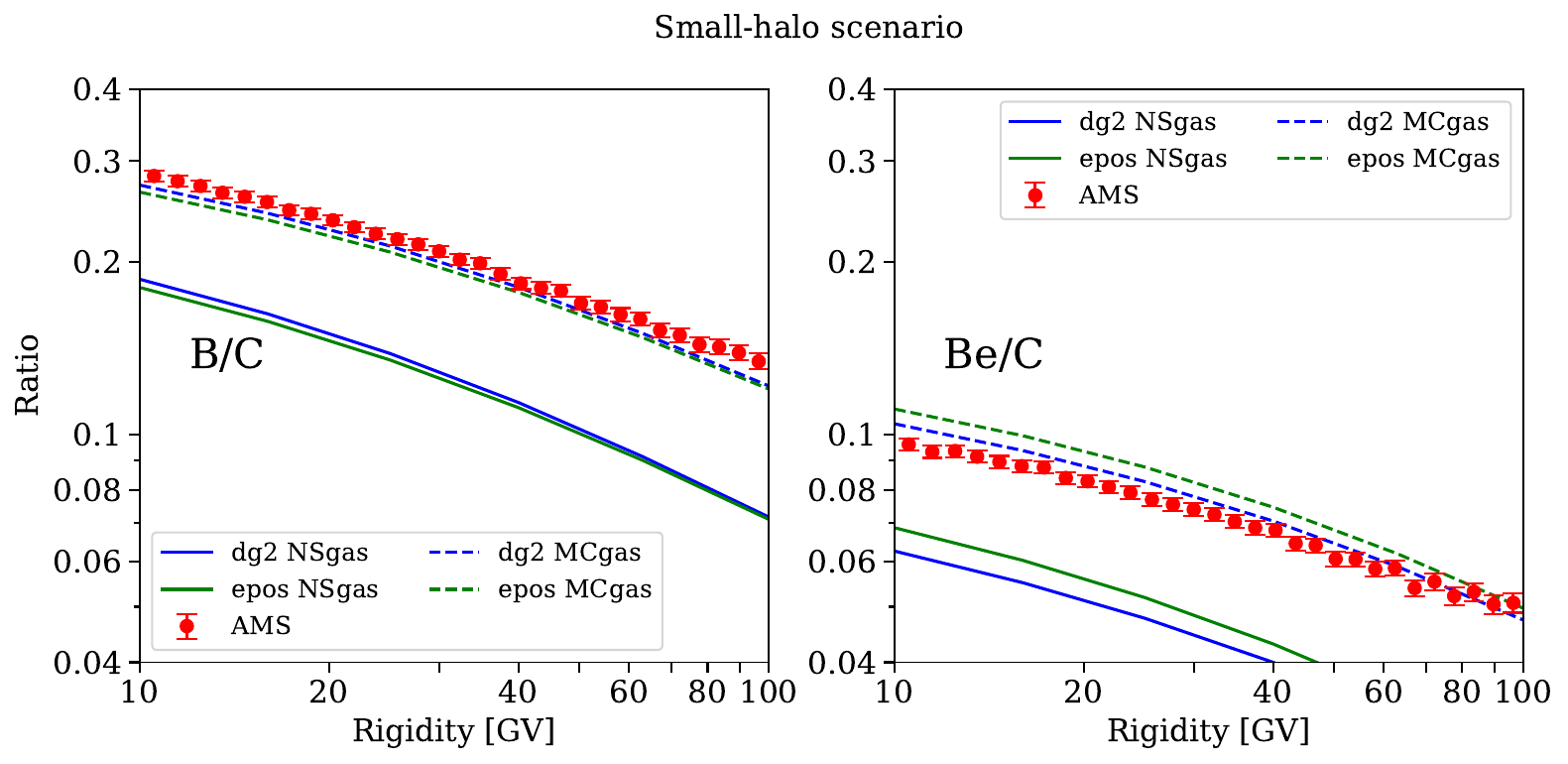}
\caption{Compare AMS data and small-halo model prediction, with different cross sections and gas templates. DRAGON2 (dg2) and EPOS-LHC (epos) cross sections are separated by blue and green colors. While Nakanishi2003/2006 (NSgas) and McMillan2017 (MCgas) gas models are separated by solid and dashed lines. The parameters are set to be: $\rm D_0 = 10^{28.4} cm^2/s, \delta = 0.6, z_H= 8 \, kpc, R_d = 16 \, kpc$. }
\label{fig:small halo fit}
\end{figure*}

Fits to the B/C and Be/C (or Be/B) ratios in the small-halo scenario have been studied extensively \citep{Mengjie_2024, Silver_2024, Korsmeier_2021, Weinrich_2020, Evoli_2020, Genolini_2019, Tomassetti_2017}. Although these studies have shown that the small-halo scenario can reproduce the data well, uncertainties in the production cross sections have been shown to remain one of the dominant limitations for constraining the halo height for this case. Here, we compare the effects of different cross-section and gas-template combinations on the predicted B/C and Be/C ratios.

Fig.~\ref{fig:small halo fit} compares the model predictions with AMS-02 measurements of the B/C and Be/C ratios between 10 and 100~GV. In this figure, all models adopt the same propagation parameters, with only the cross sections and gas templates being changed. 

The Nakanishi2003/2006 gas template predicts $\sim30\%-40\%$ fewer boron and beryllium nuclei than the McMillan2017 template. By contrast, the difference between the DRAGON2 and EPOS-LHC cross sections is much smaller: DRAGON2 predicts slightly more boron and slightly less beryllium than EPOS-LHC.

Overall, we find that the impact of the gas templates ($\sim30\%-40\%$) is significantly larger than that of the cross sections ($\lesssim3\%$) in our comparison. This mainly reflects the substantially higher gas density in the McMillan2017 model relative to the Nakanishi2003/2006 model, whereas the difference between DRAGON2 and EPOS-LHC cross sections is only marginal. These results highlight the importance of gas models derived from Galactic dynamics, which predict significantly more ISM gas than the conventional $\rm CO$ and HI survey-based models commonly used in the CR community. Nevertheless, other studies have shown that cross-section uncertainties alone can contribute up to $\sim30\%$ uncertainty in the B/C ratio, making their effect comparable in some cases. \cite{TorreLuque_2021} also noted that uncertainties in the gas distribution can affect halo-height constraints at a level similar to cross-section uncertainties. Therefore, without more precise gas templates and cross sections, robust constraints on the propagation parameters remain difficult to obtain.

\subsubsection{Giant-halo scenario}
\begin{figure*}[htbp]
\centering
\includegraphics[width=0.8\textwidth]{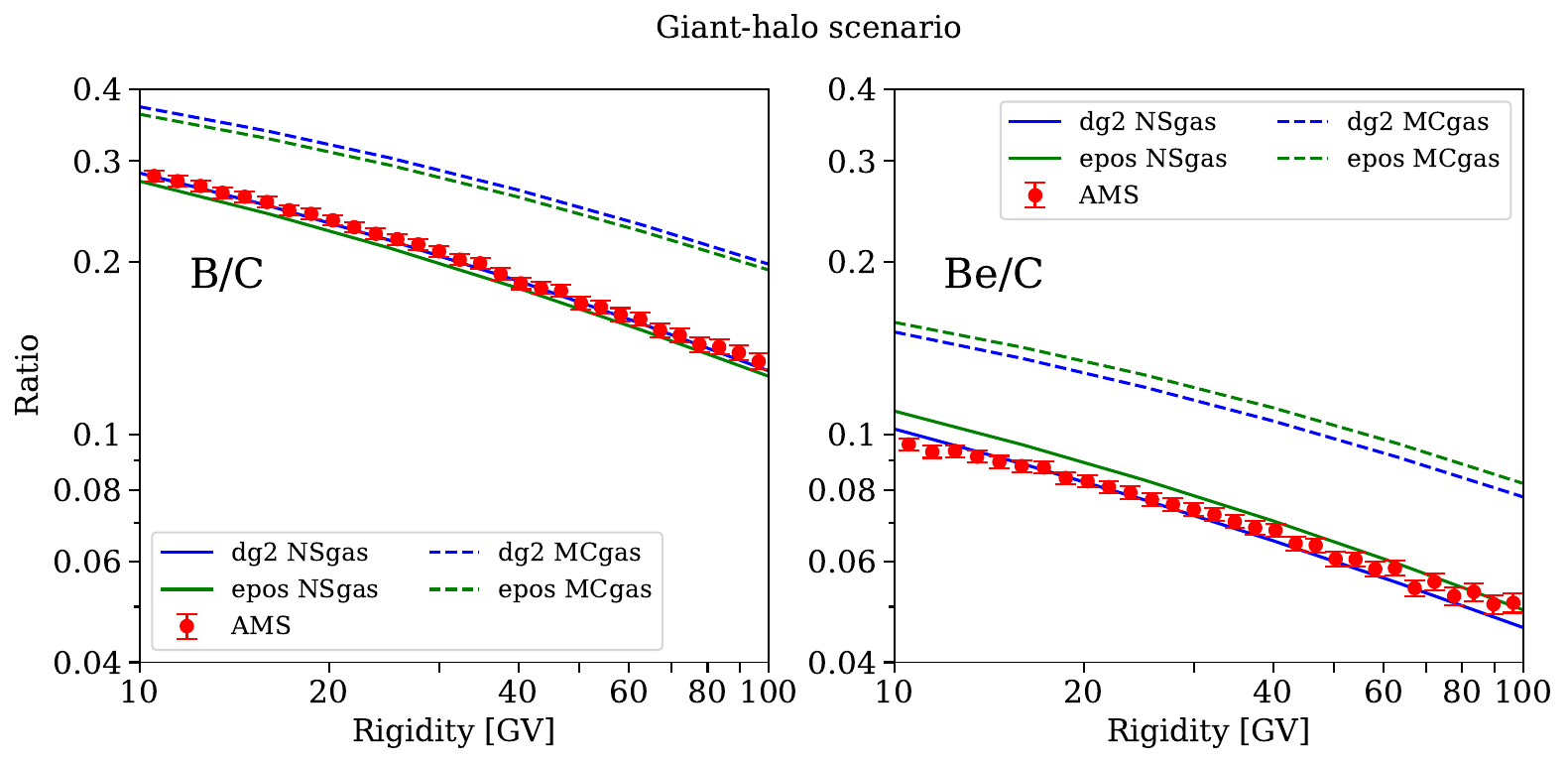}
\caption{Similar to Fig. \ref{fig:small halo fit}, but with giant-halo model prediction. The parameters are: $\rm D_0 = 10^{28.6} cm^2/s, \delta = 0.5, R_d = 20 \, kpc$. }
\label{fig:giant halo fit}
\end{figure*}

Since the giant-halo scenario is compatible with the small-halo scenario within the source region, it is expected to produce similar secondary-to-primary ratios for similar propagation parameters, as shown in Fig.~\ref{fig:giant halo fit}. Therefore, the uncertainty analysis discussed above also applies to the giant-halo scenario.

From Eq.~\ref{eq: BC_diffusion_dominated} and Eq.~\ref{eq: BC_spallation_dominated}, a purely diffusion-dominated transport regime would predict a B/C slope of $-\delta$. However, for both scenarios the diffusion coefficient index $\delta \sim 0.5$--$0.6$ leads to a B/C spectral slope of only $\sim-0.3$.  The flatter slope found here therefore suggests that CR transport is more likely spallation dominated, or in a transitional regime where diffusion and spallation contribute comparably. Consistent with this interpretation, the B/C slope obtained with the Nakanishi2003/2006 gas template is slightly steeper than that obtained with the McMillan2017 template, reflecting the effect of the lower gas density in the former model.

Compared with the small-halo scenario, the giant-halo model removes the halo height as a free parameter and therefore reduces the number of degrees of freedom, while still providing a similarly good fit to the secondary-to-primary ratios. Furthermore, the giant-halo model does not require a physical boundary at hundreds of kpc because CRs in the inner halo can reach a steady state without being affected by the outer boundary. This is a consequence of the nature of 3D random walks: once CRs propagate to sufficiently large distances, the probability of returning to the Galactic disk becomes very small. 

\section{Discussion}\label{sec: discussion}

\subsection{Observational test of the giant-halo scenario}\label{sec: disscussion.1}

In the giant-halo scenario, beyond the source radius, the CR density follows an approximately $1/r$ profile out to hundreds of kpc. Observations suggest that the CGM gas density also follows a power law, $n_{\rm gas}\propto r^{-1.2}$  \citep{Zhang_2024}. The differential rate of hadronic interactions is therefore $dN/dr\sim 4\pi r^2 r^{-1}r^{-1.2} \sim r^{-0.2}$, which declines only weakly with radius. As a result, a substantial fraction of CR--gas interactions is expected to occur in the CGM, potentially generating significant diffuse gamma-ray and neutrino emission. The observed excesses in diffuse gamma rays  \citep{LHAASO_2023,Recchia_2021,Pshirkov_2026} and neutrinos \citep{Taylor_2014, IceCube2023, Kalashev_2023} may therefore provide supporting evidence for the giant-halo scenario.

The extended age distribution in the giant halo also increases the fraction of old CRs. Since unstable nuclei (e.g., ($\rm ^{10}Be$)) are more strongly affected by radioactive decay than stable nuclei over long timescales, old unstable CRs accumulate less than old stable CRs as the halo size increases. Consequently, the normalization of unstable-to-stable ratios such as $\rm ^{10}Be/^{9}Be$ is reduced relative to the small-halo scenario. In addition, the energy dependence of the ratio becomes flatter because the growing decay timescale with Lorentz factor has a weaker impact when CRs are distributed in a broader age range \citep{Lipari_2022}. This provides another potential test of the giant-halo scenario.

\subsection{How uncertainties affect $ {}^{10}Be/{}^{9}Be$ ratio}

Although the total beryllium production cross sections are similar in the DRAGON2 and EPOS-LHC models, the predicted production channels for $\rm {}^{10}Be$ differ substantially. It remains uncertain whether $\rm {}^{10}Be$ is produced predominantly by primary CRs ($\rm {}^{12}C$ and $\rm {}^{16}O$) or by secondary CRs (N, $\rm {}^{13}C$, and B). While primary species have larger fluxes, their $\rm {}^{10}Be$ production cross sections are smaller; secondary species show the opposite behavior. A rough estimate of the relative contributions can be obtained from the product of the steady-state flux and the production cross section.

At 10--100 GV, the flux of secondary CRs is approximately $10\%-30\%$ of that of primary CRs. In contrast, the total $\rm {}^{10}Be$ production cross section summed over secondary CRs is about 3 times that summed over primary CRs in DRAGON2, and about 20 times in EPOS-LHC. As a result, EPOS-LHC predicts that $\rm {}^{10}Be$ production is dominated by secondary CRs, whereas DRAGON2 predicts comparable contributions from primary and secondary CRs. This demonstrates that cross-section uncertainties affect not only the overall secondary CR flux but also the relative importance of isotope production channels.

Gas-template uncertainties also influence the $\rm {}^{10}Be/{}^{9}Be$ ratio. A higher gas density is expected to increase the ratio because the two isotopes probe different spatial scales. The $\rm {}^{10}Be$ flux observed at Earth originates from a local region characterized by the diffusion length before decay, $R_{decay}\sim\sqrt{D\tau_d}$, whereas $\rm {}^{9}Be$ samples a larger region characterized by the diffusion length before spallation, $R_{spallation}\sim\sqrt{D\tau_s}$. Increasing the gas density reduces $R_{spallation}$ and thus the effective production volume of $\rm {}^{9}Be$, while having little impact on the local origin of $\rm {}^{10}Be$. Consequently, the $\rm {}^{10}Be/{}^{9}Be$ ratio increases.

Fig.~\ref{fig:Be10/Be9} compares the predicted $\rm {}^{10}Be/{}^{9}Be$ ratios for different combinations of cross-section models and gas templates, using the same propagation parameters as in Figs~\ref{fig:small halo fit} and \ref{fig:giant halo fit}. The DRAGON2 cross sections predict significantly higher $\rm {}^{10}Be/{}^{9}Be$ ratios than the EPOS-LHC cross-sections, while the McMillan2017 gas template predicts higher ratios than the Nakanishi2003/2006 template. These trends are consistent with the qualitative arguments above. Overall, uncertainties in both nuclear cross sections and gas distributions have a strong impact on $\rm {}^{10}Be$ production and therefore on the inferred CR confinement time in the Galaxy.

\subsection{Limitations of our study}

The present study adopts a simplified transport model based on isotropic and spatially uniform diffusion coefficient. In reality, CR transport may be anisotropic, spatially dependent, and advection may also play a relevant role in certain regions. However, given our limited present knowledge of the Galactic magnetic-field structures, turbulence in the CGM, and large-scale Galactic winds, the model adopted should be considered as a reduced simplified model aimed at encapsulating the key features of CR propagation beyond the Galactic disk.

Our results identify that the source size plays a key scale in governing the transition between quasi-one-dimensional and three-dimensional transport due to geometry effect. In more realistic scenarios, however, additional characteristic scales are likely to arise. For example, transitions between nature of the magnetic turbulence in the Galactic disk and in the CGM, or between diffusion-dominated and advection-dominated propagation, may introduce new length scales that affect both the CR age distribution and their spatial distribution in the CGM. Exploring the interplay among these scales, warranted by the improvement in observational results expected in the coming years, will require more sophisticated models, potentially incorporating anisotropic diffusion, spatially varying transport coefficients, and advective transport.

\section{Conclusion}
\label{sec:conclusion}

The transport of CRs in a giant CGM halo has been extensively studied, particularly its similarities to, and differences from, the conventional small-halo scenario. In the giant-halo framework, the source size is a key parameter that plays a role mathematically analogous to the halo height in the conventional small-halo model.

Within the source region, the CR spatial distribution closely resembles that of the small-halo scenario, with transport effectively reduced to 1D diffusion along the z-direction. On these scales, the age distribution at Earth follows $dn/dt \propto t^{-1/2}$, matching the behavior of the 1D small-halo model. Consequently, the giant-halo scenario can reproduce the observational successes of the small-halo framework. The main uncertainties arise from nuclear cross sections and gas templates, which contribute at comparable levels and still significantly affect the predicted CR fluxes and ratios.

Beyond the source region, CR transport becomes fully 3D, producing a 1/r radial tail that can extend to hundreds of kpc. The age distribution correspondingly changes to $dn/dt \propto t^{-3/2}$, in contrast to the exponential cutoff of the small-halo scenario. This broader age distribution can noticeably modify CR fluxes and composition ratios, particularly for unstable secondary species.

Recent measurements by AMS-02 and DAMPE have determined CR fluxes and composition ratios with unprecedented precision, providing stringent tests of propagation models. In particular, $\rm ^{10}Be$ serves as a standard chronometer because its half-life ($\sim 1.39\times 10^6$ yr) is comparable to the Galactic CR confinement time. The forthcoming high-precision measurements of the $\rm ^{10}Be / ^9 Be$ ratio from AMS-02 will be essential to test the giant-halo scenario.


%

\vspace{5mm}




\appendix

\section{$^{10}\mathrm{B\lowercase{e}}/^{9}\mathrm{B\lowercase{e}}$ Ratio in  small \& giant-halo scenarios}

The $\rm {}^{10}Be/{}^{9}Be$ ratio is a key observable for constraining the CR age distribution. A qualitative discussion was presented in Sec. \ref{sec: disscussion.1}. Here, we show model predictions for the $\rm {}^{10}Be/{}^{9}Be$ ratio using the same parameter sets as in Figs. \ref{fig:small halo fit} and \ref{fig:giant halo fit}. Direct quantitative comparison between the small- and giant-halo scenarios, however, is not straightforward because the two models have different number of parameters.

\begin{figure*}[htbp]
    \centering
    
    \begin{minipage}[t]{0.48\linewidth}
        \centering
        \includegraphics[width=\linewidth]{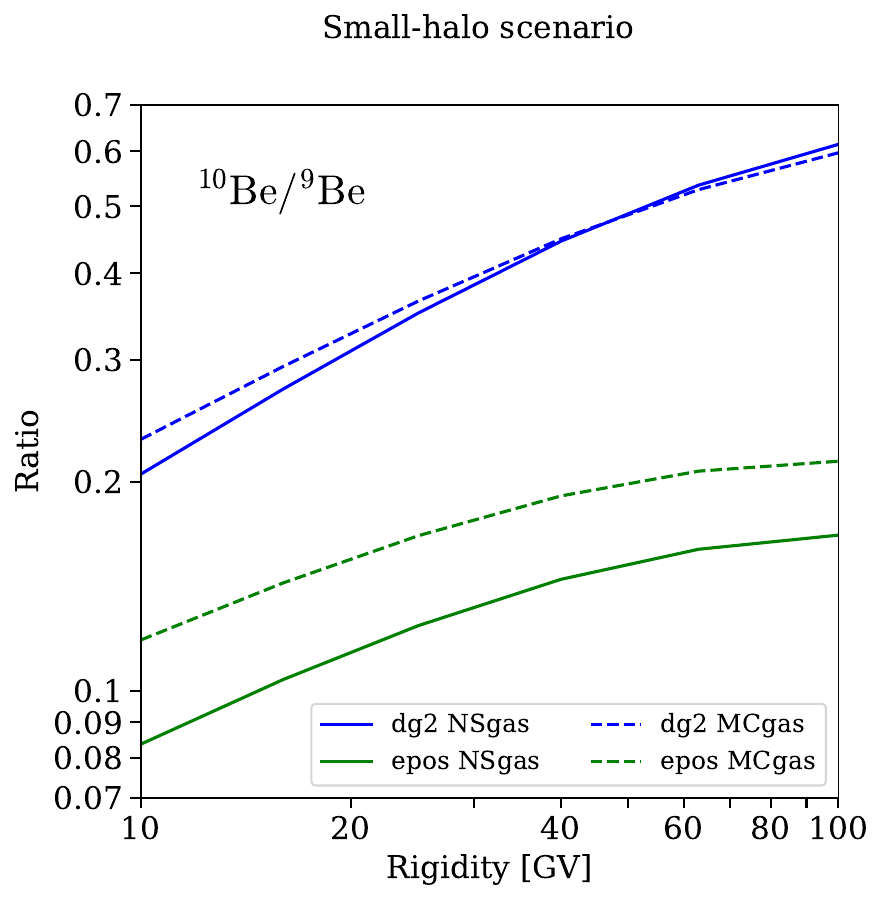}
    \end{minipage}
    \hfill
    \begin{minipage}[t]{0.48\linewidth}
        \centering
        \includegraphics[width=\linewidth]{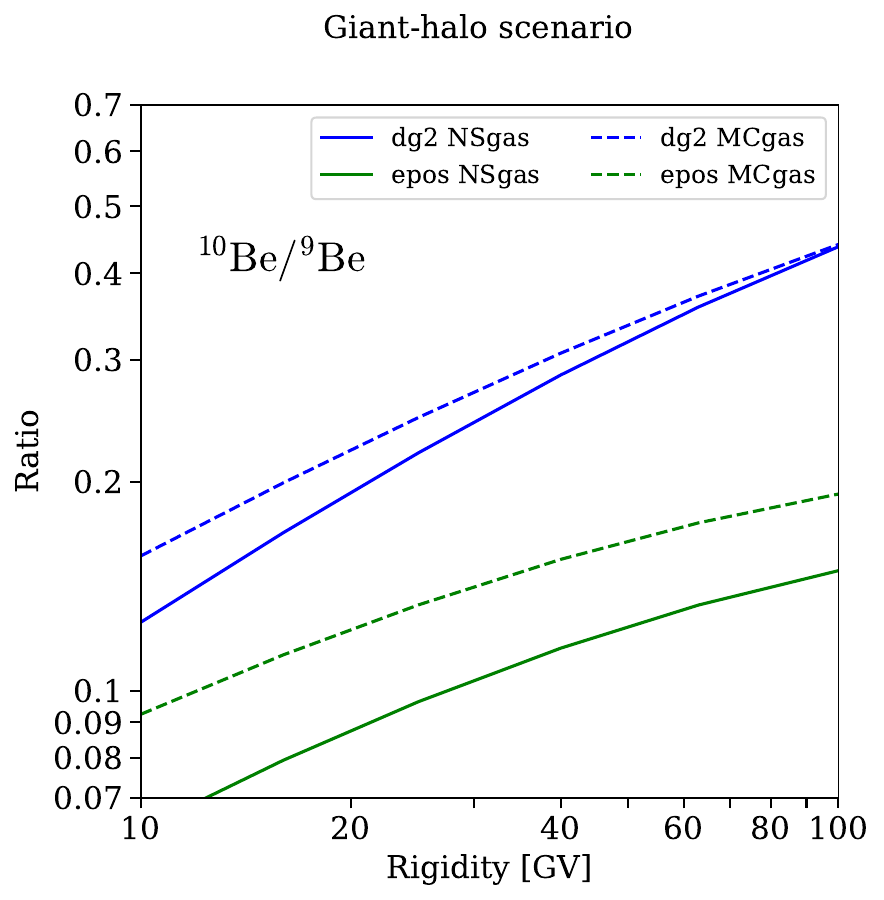}
    \end{minipage}

    \caption{Predicted $\rm {}^{10}Be/{}^{9}Be$ ratio for both small-halo(left) and giant-halo(right) scenarios with different cross section and gas template combinations. The parameters are set to be: $\rm D_0 = 10^{28.4} cm^2/s, \delta = 0.6, z_H= 8.0 \, kpc, R_d = 16.0 \, kpc$ for the small-halo scenario and $\rm D_0 = 10^{28.6} cm^2/s, \delta = 0.5, R_d = 20.0 \, kpc$ for the giant-halo scenario. }
    \label{fig:Be10/Be9}
\end{figure*}

\section{Analytical solution of 3D diffusion}\label{sec: appendix_1}

Analytical solution in very simple cases can provide intuitive understanding of distribution and secondary-to-primary ratio of CRs. However, to obtain such results, simplifications must first be made. We here consider a reduced scenario in which a single stable primary species ($n_1$) produces a single secondary species ($n_2$) within a uniformly distributed gas. The transport of these CRs is then described by the following equation:

\begin{equation}\label{eq:diffusion_equation_1}
\left \{
\begin{aligned}
    & \frac{\partial n_1}{\partial t} = D\nabla^2 n_1 - \frac{n_1}{\tau_1} + q\\
    & \frac{\partial n_2}{\partial t} = D\nabla^2 n_2 - \frac{n_2}{\tau_2} + \frac{n_p}{\tau_{1\to 2}}
\end{aligned}
\right.
\end{equation}
where $\tau_1,\tau_2$, and $ \tau_{1\to 2}$ are spallation timescales for CR species 1, 2 and from 1 to produce 2. 

Further simplification is to assume CRs are injected from a point source, and a spherically symmetry absorbing boundary is at $R_H$. The escape timescale is defined as $t_{escape} = R_H^2/D$. If $\tau_{1,2} \ll t_{escape}$, CRs undergo spallation before reaching the boundary, and the steady-state solution is:
\begin{equation} \label{eq:solution_spallatin_dominated}
 \left\{
\begin{aligned}
    & n_1(r) = \frac{q}{4\pi D r} \exp(-\frac{r}{\sqrt{D\tau_1}}) \\
    & n_2(r) = \frac{q}{4\pi Dr} \frac{\tau_1\tau_2}{\tau_{1\to 2}(\tau_2-\tau_1)}\times \Big [ \exp(-\frac{r}{\sqrt{D\tau_2}}) - \exp(-\frac{r}{\sqrt{D\tau_1}}) \Big ] 
\end{aligned}
\right.
\end{equation}
It shows the distribution for primary CR is a power-law($r^{-1}$) with an exponential cutoff, and the cutoff distance is the diffusion length before CR spallation. The distribution for secondary CR is more complicated, dictated by the difference between two spallation timescales.

For diffusion dominated transport, $\tau_{1,2} \gg t_{escape}$, we can ignore the spallation terms($n_1 / \tau_1, n_2 / \tau_2$) in Eq. \eqref{eq:diffusion_equation_1} and the solutions are:

\begin{equation} \label{eq:solution_diffusion_dominated}
 \left\{
\begin{aligned}
    & n_1(r) = \frac{q}{4\pi D} (\frac{1}{r} - \frac{1}{R_H})  \\
    & n_2(r) = \frac{q}{4\pi D^2 \tau}\frac{(R_H-r)(2R_H-r)}{6R_H}
\end{aligned}
\right.
\end{equation}
The distribution for primary CR is still a power-law and sharply cut by the absorbing boundary. For secondary CRs, the distribution is almost flat when $r << R_H$ and then sharply cut at $r = R_H$.

The secondary-to-primary ratio is
\begin{equation}\label{eq: BC_spallation_dominated}
    \frac{n_2}{n_1} = \frac{\tau_2\tau_1}{\tau_{1\to 2}(\tau_2 - \tau_1)}  \Big [ \exp(\frac{r}{\sqrt{D\tau_1}}-\frac{r}{\sqrt{D\tau_2}}) - 1 \Big] \sim \frac{1}{D^{1/2}}
\end{equation}\label{eq: BC_diffusion_dominated}
when spallation dominates, and
\begin{equation}
    \frac{n_2}{n_1} = \frac{(2R_H-r)r}{6D\tau_{1\to2}} \sim \frac{R_H}{D}
\end{equation}
when diffusion dominates.

The approximation in Eq. (\ref{eq: BC_spallation_dominated}) assumes that $\tau_2 = (1+\epsilon)\tau_1$ where $\epsilon$ is a small number. These results show that the secondary-to-primary ratio scales as $R_H/D$ in the diffusion-dominated regime, consistent with the small-halo scenario. In contrast, the dependence becomes flatter in the spallation-dominated regime, approximately $\propto D^{-1/2}$.

If the secondary CR is unstable, we can rewrite
\begin{equation}
    \frac{1}{\tau_2} = \frac{1}{\tau_{s,2}} + \frac{1}{\gamma\tau_{d,2}}
\end{equation}
and find that it evolves with rigidity. One typical example is $\rm {}^{10}Be$, whose decay timescale is around $\tau_d = 1.4\times 10^6$ years. Its spallation timescale in ISM is $\rm \tau_s = 5.4\times 10^6 (1 \, cm^{-3} / n_H) $ year. For non-relativistic $\rm {}^{10}Be$, its loss is dominated by decay. But for relativistic $\rm {}^{10}Be$, its loss is dominated by spallation. Such transition makes $\rm {}^{10}Be$ a unique probe for the CR residence time in the galaxy\citep{Evoli_2020}.


\bibliography{sample631}{}

@ARTICLE{Cautun_2020,
       author = {{Cautun}, Marius and {Ben{\'\i}tez-Llambay}, Alejandro and {Deason}, Alis J. and {Frenk}, Carlos S. and {Fattahi}, Azadeh and {G{\'o}mez}, Facundo A. and {Grand}, Robert J.~J. and {Oman}, Kyle A. and {Navarro}, Julio F. and {Simpson}, Christine M.},
        title = "{The milky way total mass profile as inferred from Gaia DR2}",
      journal = {mnras},
         year = 2020,
        month = may,
       volume = {494},
       number = {3},
        pages = {4291-4313},
          doi = {10.1093/mnras/staa1017},
archivePrefix = {arXiv},
       eprint = {1911.04557},
 primaryClass = {astro-ph.GA},
       adsurl = {https://ui.adsabs.harvard.edu/abs/2020MNRAS.494.4291C}
}

@ARTICLE{Zhang_2024,
       author = {{Zhang}, Yi and {Comparat}, Johan and {Ponti}, Gabriele and {Merloni}, Andrea and {Nandra}, Kirpal and {Haberl}, Frank and {Locatelli}, Nicola and {Zhang}, Xiaoyuan and {Sanders}, Jeremy and {Zheng}, Xueying and {Liu}, Ang and {Popesso}, Paola and {Liu}, Teng and {Truong}, Nhut and {Pillepich}, Annalisa and {Predehl}, Peter and {Salvato}, Mara and {Shreeram}, Soumya and {Yeung}, Michael C.~H. and {Ni}, Qingling},
        title = "{The hot circumgalactic medium in the eROSITA All-Sky Survey: I. X-ray surface brightness profiles}",
      journal = {aap},
         year = 2024,
        month = oct,
       volume = {690},
          eid = {A267},
        pages = {A267},
          doi = {10.1051/0004-6361/202449412},
archivePrefix = {arXiv},
       eprint = {2401.17308},
 primaryClass = {astro-ph.GA},
       adsurl = {https://ui.adsabs.harvard.edu/abs/2024A&A...690A.267Z}
}

@ARTICLE{Maller_2004,
       author = {{Maller}, Ariyeh H. and {Bullock}, James S.},
        title = "{Multiphase galaxy formation: high-velocity clouds and the missing baryon problem}",
      journal = {mnras},
         year = 2004,
        month = dec,
       volume = {355},
       number = {3},
        pages = {694-712},
          doi = {10.1111/j.1365-2966.2004.08349.x},
archivePrefix = {arXiv},
       eprint = {astro-ph/0406632},
 primaryClass = {astro-ph},
       adsurl = {https://ui.adsabs.harvard.edu/abs/2004MNRAS.355..694M}
}

@ARTICLE{Evoli_2020,
       author = {{Evoli}, Carmelo and {Morlino}, Giovanni and {Blasi}, Pasquale and {Aloisio}, Roberto},
        title = "{AMS-02 beryllium data and its implication for cosmic ray transport}",
      journal = {prd},
         year = 2020,
        month = jan,
       volume = {101},
       number = {2},
          eid = {023013},
        pages = {023013},
          doi = {10.1103/PhysRevD.101.023013},
archivePrefix = {arXiv},
       eprint = {1910.04113},
 primaryClass = {astro-ph.HE},
       adsurl = {https://ui.adsabs.harvard.edu/abs/2020PhRvD.101b3013E}
}

@ARTICLE{Genolini_2019,
       author = {{G{\'e}nolini}, Y. and {Boudaud}, M. and {Batista}, P. -I. and {Caroff}, S. and {Derome}, L. and {Lavalle}, J. and {Marcowith}, A. and {Maurin}, D. and {Poireau}, V. and {Poulin}, V. and {Rosier}, S. and {Salati}, P. and {Serpico}, P.~D. and {Vecchi}, M.},
        title = "{Cosmic-ray transport from AMS-02 boron to carbon ratio data: Benchmark models and interpretation}",
      journal = {prd},
         year = 2019,
        month = jun,
       volume = {99},
       number = {12},
          eid = {123028},
        pages = {123028},
          doi = {10.1103/PhysRevD.99.123028},
archivePrefix = {arXiv},
       eprint = {1904.08917},
 primaryClass = {astro-ph.HE},
       adsurl = {https://ui.adsabs.harvard.edu/abs/2019PhRvD..99l3028G}
}

@ARTICLE{Evoli_2019,
       author = {{Evoli}, Carmelo and {Aloisio}, Roberto and {Blasi}, Pasquale},
        title = "{Galactic cosmic rays after the AMS-02 observations}",
      journal = {prd},
         year = 2019,
        month = may,
       volume = {99},
       number = {10},
          eid = {103023},
        pages = {103023},
          doi = {10.1103/PhysRevD.99.103023},
archivePrefix = {arXiv},
       eprint = {1904.10220},
 primaryClass = {astro-ph.HE},
       adsurl = {https://ui.adsabs.harvard.edu/abs/2019PhRvD..99j3023E}
}

@software{Ulrich_2021,
       author = {{Ulrich}, Ralf and {Pierog}, Tanguy and {Baus}, Colin},
        title = "{Cosmic Ray Monte Carlo Package, CRMC}",
         year = 2021,
        month = aug,
          eid = {10.5281/zenodo.5270381},
          doi = {10.5281/zenodo.5270381},
      version = {2.0.1},
    publisher = {Zenodo},
       adsurl = {https://ui.adsabs.harvard.edu/abs/2021zndo...5270381U}
}

@ARTICLE{Tumlinson_2017,
       author = {{Tumlinson}, Jason and {Peeples}, Molly S. and {Werk}, Jessica K.},
        title = "{The Circumgalactic Medium}",
      journal = {araa},
         year = 2017,
        month = aug,
       volume = {55},
       number = {1},
        pages = {389-432},
          doi = {10.1146/annurev-astro-091916-055240},
archivePrefix = {arXiv},
       eprint = {1709.09180},
 primaryClass = {astro-ph.GA},
       adsurl = {https://ui.adsabs.harvard.edu/abs/2017ARA&A..55..389T}
}

@article{Bregman_2022,
doi = {10.3847/1538-4357/ac51de},
url = {https://dx.doi.org/10.3847/1538-4357/ac51de},
year = {2022},
month = {mar},
publisher = {The American Astronomical Society},
volume = {928},
number = {1},
pages = {14},
author = {Bregman, Joel N. and Hodges-Kluck, Edmund and Qu, Zhijie and Pratt, Cameron and Li, Jiang-Tao and Yun, Yansong},
title = {Hot Extended Galaxy Halos around Local L* Galaxies from Sunyaev–Zeldovich Measurements},
journal = {apj},
}

@ARTICLE{Cox_2005,
       author = {{Cox}, Donald P.},
        title = "{The Three-Phase Interstellar Medium Revisited}",
      journal = {araa},
         year = 2005,
        month = sep,
       volume = {43},
       number = {1},
        pages = {337-385},
          doi = {10.1146/annurev.astro.43.072103.150615},
       adsurl = {https://ui.adsabs.harvard.edu/abs/2005ARA&A..43..337C}
}

@article{AMS_2021,
title = {The Alpha Magnetic Spectrometer (AMS) on the international space station: Part II — Results from the first seven years},
journal = {Physics Reports},
volume = {894},
pages = {1-116},
year = {2021},
note = {The Alpha Magnetic Spectrometer (AMS) on the International Space Station: Part II - Results from the First Seven Years},
issn = {0370-1573},
doi = {https://doi.org/10.1016/j.physrep.2020.09.003},
url = {https://www.sciencedirect.com/science/article/pii/S0370157320303434},
author = {M. Aguilar and L. {Ali Cavasonza} and G. Ambrosi and L. Arruda and N. Attig and F. Barao and L. Barrin and A. Bartoloni and S. {Başeğmez-du Pree} and J. Bates and R. Battiston and M. Behlmann and B. Beischer and J. Berdugo and B. Bertucci and V. Bindi and W. {de Boer} and K. Bollweg and B. Borgia and M.J. Boschini and M. Bourquin and E.F. Bueno and J. Burger and W.J. Burger and S. Burmeister and X.D. Cai and M. Capell and J. Casaus and G. Castellini and F. Cervelli and Y.H. Chang and G.M. Chen and H.S. Chen and Y. Chen and L. Cheng and H.Y. Chou and S. Chouridou and V. Choutko and C.H. Chung and C. Clark and G. Coignet and C. Consolandi and A. Contin and C. Corti and Z. Cui and K. Dadzie and Y.M. Dai and C. Delgado and S. {Della Torre} and M.B. Demirköz and L. Derome and S. {Di Falco} and V. {Di Felice} and C. Díaz and F. Dimiccoli and P. {von Doetinchem} and F. Dong and F. Donnini and M. Duranti and A. Egorov and A. Eline and J. Feng and E. Fiandrini and P. Fisher and V. Formato and C. Freeman and Y. Galaktionov and C. Gámez and R.J. García-López and C. Gargiulo and H. Gast and I. Gebauer and M. Gervasi and F. Giovacchini and D.M. Gómez-Coral and J. Gong and C. Goy and V. Grabski and D. Grandi and M. Graziani and K.H. Guo and S. Haino and K.C. Han and R.K. Hashmani and Z.H. He and B. Heber and T.H. Hsieh and J.Y. Hu and Z.C. Huang and W. Hungerford and M. Incagli and W.Y. Jang and Yi Jia and H. Jinchi and K. Kanishev and B. Khiali and G.N. Kim and Th. Kirn and M. Konyushikhin and O. Kounina and A. Kounine and V. Koutsenko and A. Kuhlman and A. Kulemzin and G. {La Vacca} and E. Laudi and G. Laurenti and I. Lazzizzera and A. Lebedev and H.T. Lee and S.C. Lee and C. Leluc and J.Q. Li and M. Li and Q. Li and S. Li and T.X. Li and Z.H. Li and C. Light and C.H. Lin and T. Lippert and Z. Liu and S.Q. Lu and Y.S. Lu and K. Luebelsmeyer and J.Z. Luo and S.S. Lyu and F. Machate and C. Mañá and J. Marín and J. Marquardt and T. Martin and G. Martínez and N. Masi and D. Maurin and A. Menchaca-Rocha and Q. Meng and D.C. Mo and M. Molero and P. Mott and L. Mussolin and J.Q. Ni and N. Nikonov and F. Nozzoli and A. Oliva and M. Orcinha and M. Palermo and F. Palmonari and M. Paniccia and A. Pashnin and M. Pauluzzi and S. Pensotti and H.D. Phan and V. Plyaskin and M. Pohl and S. Porter and X.M. Qi and X. Qin and Z.Y. Qu and L. Quadrani and P.G. Rancoita and D. Rapin and A. {Reina Conde} and S. Rosier-Lees and A. Rozhkov and D. Rozza and R. Sagdeev and S. Schael and S.M. Schmidt and A. {Schulz von Dratzig} and G. Schwering and E.S. Seo and B.S. Shan and J.Y. Shi and T. Siedenburg and C. Solano and J.W. Song and R. Sonnabend and Q. Sun and Z.T. Sun and M. Tacconi and X.W. Tang and Z.C. Tang and J. Tian and Samuel C.C. Ting and S.M. Ting and N. Tomassetti and J. Torsti and C. Tüysüz and T. Urban and I. Usoskin and V. Vagelli and R. Vainio and E. Valente and E. Valtonen and M. {Vázquez Acosta} and M. Vecchi and M. Velasco and J.P. Vialle and L.Q. Wang and N.H. Wang and Q.L. Wang and S. Wang and X. Wang and Z.X. Wang and J. Wei and Z.L. Weng and H. Wu and R.Q. Xiong and W. Xu and Q. Yan and Y. Yang and H. Yi and Y.J. Yu and Z.Q. Yu and M. Zannoni and C. Zhang and F. Zhang and F.Z. Zhang and J.H. Zhang and Z. Zhang and F. Zhao and Z.M. Zheng and H.L. Zhuang and V. Zhukov and A. Zichichi and N. Zimmermann and P. Zuccon},
}

@ARTICLE{Blasi_2012,
       author = {{Blasi}, Pasquale and {Amato}, Elena},
        title = "{Diffusive propagation of cosmic rays from supernova remnants in the Galaxy. I: spectrum and chemical composition}",
      journal = {jcap},
         year = 2012,
        month = jan,
       volume = {2012},
       number = {1},
          eid = {010},
        pages = {010},
          doi = {10.1088/1475-7516/2012/01/010},
archivePrefix = {arXiv},
       eprint = {1105.4521},
 primaryClass = {astro-ph.HE},
       adsurl = {https://ui.adsabs.harvard.edu/abs/2012JCAP...01..010B}
}

@ARTICLE{Taylor_2014,
       author = {{Taylor}, Andrew M. and {Gabici}, Stefano and {Aharonian}, Felix},
        title = "{Galactic halo origin of the neutrinos detected by IceCube}",
      journal = {prd},
         year = 2014,
        month = may,
       volume = {89},
       number = {10},
          eid = {103003},
        pages = {103003},
          doi = {10.1103/PhysRevD.89.103003},
archivePrefix = {arXiv},
       eprint = {1403.3206},
 primaryClass = {astro-ph.HE},
       adsurl = {https://ui.adsabs.harvard.edu/abs/2014PhRvD..89j3003T}
}

@ARTICLE{Tomassetti_2017,
       author = {{Tomassetti}, Nicola},
        title = "{Solar and nuclear physics uncertainties in cosmic-ray propagation}",
      journal = {prd},
         year = 2017,
        month = nov,
       volume = {96},
       number = {10},
          eid = {103005},
        pages = {103005},
          doi = {10.1103/PhysRevD.96.103005},
archivePrefix = {arXiv},
       eprint = {1707.06917},
 primaryClass = {astro-ph.HE},
       adsurl = {https://ui.adsabs.harvard.edu/abs/2017PhRvD..96j3005T}
}

@ARTICLE{Strong_2007,
       author = {{Strong}, Andrew W. and {Moskalenko}, Igor V. and {Ptuskin}, Vladimir S.},
        title = "{Cosmic-Ray Propagation and Interactions in the Galaxy}",
      journal = {Annual Review of Nuclear and Particle Science},
         year = 2007,
        month = nov,
       volume = {57},
       number = {1},
        pages = {285-327},
          doi = {10.1146/annurev.nucl.57.090506.123011},
archivePrefix = {arXiv},
       eprint = {astro-ph/0701517},
 primaryClass = {astro-ph},
       adsurl = {https://ui.adsabs.harvard.edu/abs/2007ARNPS..57..285S}
}

@ARTICLE{Evoli_2018,
       author = {{Evoli}, Carmelo and {Gaggero}, Daniele and {Vittino}, Andrea and {Di Mauro}, Mattia and {Grasso}, Dario and {Mazziotta}, Mario Nicola},
        title = "{Cosmic-ray propagation with DRAGON2: II. Nuclear interactions with the interstellar gas}",
      journal = {jcap},
         year = 2018,
        month = jul,
       volume = {2018},
       number = {7},
          eid = {006},
        pages = {006},
          doi = {10.1088/1475-7516/2018/07/006},
archivePrefix = {arXiv},
       eprint = {1711.09616},
 primaryClass = {astro-ph.HE},
       adsurl = {https://ui.adsabs.harvard.edu/abs/2018JCAP...07..006E}
}

@ARTICLE{Dragon_I,
       author = {{Evoli}, Carmelo and {Gaggero}, Daniele and {Vittino}, Andrea and {Di Bernardo}, Giuseppe and {Di Mauro}, Mattia and {Ligorini}, Arianna and {Ullio}, Piero and {Grasso}, Dario},
        title = "{Cosmic-ray propagation with DRAGON2: I. numerical solver and astrophysical ingredients}",
      journal = {jcap},
         year = 2017,
        month = feb,
       volume = {2017},
       number = {2},
          eid = {015},
        pages = {015},
          doi = {10.1088/1475-7516/2017/02/015},
archivePrefix = {arXiv},
       eprint = {1607.07886},
 primaryClass = {astro-ph.HE},
       adsurl = {https://ui.adsabs.harvard.edu/abs/2017JCAP...02..015E}
}

@ARTICLE{Klypin_2002,
       author = {{Klypin}, Anatoly and {Zhao}, HongSheng and {Somerville}, Rachel S.},
        title = "{{\ensuremath{\Lambda}}CDM-based Models for the Milky Way and M31. I. Dynamical Models}",
      journal = {apj},
         year = 2002,
        month = jul,
       volume = {573},
       number = {2},
        pages = {597-613},
          doi = {10.1086/340656},
archivePrefix = {arXiv},
       eprint = {astro-ph/0110390},
 primaryClass = {astro-ph},
       adsurl = {https://ui.adsabs.harvard.edu/abs/2002ApJ...573..597K}
}

@ARTICLE{Bhattacharjee_2014,
       author = {{Bhattacharjee}, Pijushpani and {Chaudhury}, Soumini and {Kundu}, Susmita},
        title = "{Rotation Curve of the Milky Way out to \raisebox{-0.5ex}\textasciitilde200 kpc}",
      journal = {apj},
         year = 2014,
        month = apr,
       volume = {785},
       number = {1},
          eid = {63},
        pages = {63},
          doi = {10.1088/0004-637X/785/1/63},
archivePrefix = {arXiv},
       eprint = {1310.2659},
 primaryClass = {astro-ph.GA},
       adsurl = {https://ui.adsabs.harvard.edu/abs/2014ApJ...785...63B}
}

@ARTICLE{Faerman_2017,
       author = {{Faerman}, Yakov and {Sternberg}, Amiel and {McKee}, Christopher F.},
        title = "{Massive Warm/Hot Galaxy Coronae as Probed by UV/X-Ray Oxygen Absorption and Emission. I. Basic Model}",
      journal = {apj},
         year = 2017,
        month = jan,
       volume = {835},
       number = {1},
          eid = {52},
        pages = {52},
          doi = {10.3847/1538-4357/835/1/52},
archivePrefix = {arXiv},
       eprint = {1602.00689},
 primaryClass = {astro-ph.GA},
       adsurl = {https://ui.adsabs.harvard.edu/abs/2017ApJ...835...52F}
}

@ARTICLE{Tepper_2015,
       author = {{Tepper-Garc{\'\i}a}, Thor and {Bland-Hawthorn}, Joss and {Sutherland}, Ralph S.},
        title = "{The Magellanic Stream: Break-up and Accretion onto the Hot Galactic Corona}",
      journal = {apj},
         year = 2015,
        month = nov,
       volume = {813},
       number = {2},
          eid = {94},
        pages = {94},
          doi = {10.1088/0004-637X/813/2/94},
archivePrefix = {arXiv},
       eprint = {1505.01587},
 primaryClass = {astro-ph.GA},
       adsurl = {https://ui.adsabs.harvard.edu/abs/2015ApJ...813...94T}
}

@ARTICLE{Hopkins_2022,
       author = {{Hopkins}, Philip F. and {Butsky}, Iryna S. and {Panopoulou}, Georgia V. and {Ji}, Suoqing and {Quataert}, Eliot and {Faucher-Gigu{\`e}re}, Claude-Andr{\'e} and {Kere{\v{s}}}, Du{\v{s}}an},
        title = "{First predicted cosmic ray spectra, primary-to-secondary ratios, and ionization rates from MHD galaxy formation simulations}",
      journal = {mnras},
         year = 2022,
        month = nov,
       volume = {516},
       number = {3},
        pages = {3470-3514},
          doi = {10.1093/mnras/stac1791},
archivePrefix = {arXiv},
       eprint = {2109.09762},
 primaryClass = {astro-ph.HE},
       adsurl = {https://ui.adsabs.harvard.edu/abs/2022MNRAS.516.3470H}
}

@ARTICLE{Mori_2009,
       author = {{Mori}, Masaki},
        title = "{Nuclear enhancement factor in calculation of Galactic diffuse gamma-rays: A new estimate with DPMJET-3}",
      journal = {Astroparticle Physics},
         year = 2009,
        month = jun,
       volume = {31},
       number = {5},
        pages = {341-343},
          doi = {10.1016/j.astropartphys.2009.03.004},
archivePrefix = {arXiv},
       eprint = {0903.3260},
 primaryClass = {astro-ph.HE},
       adsurl = {https://ui.adsabs.harvard.edu/abs/2009APh....31..341M}
}

@ARTICLE{Ferriere_2007,
       author = {{Ferri{\`e}re}, K. and {Gillard}, W. and {Jean}, P.},
        title = "{Spatial distribution of interstellar gas in the innermost 3 kpc of our galaxy}",
      journal = {aap},
         year = 2007,
        month = may,
       volume = {467},
       number = {2},
        pages = {611-627},
          doi = {10.1051/0004-6361:20066992},
archivePrefix = {arXiv},
       eprint = {astro-ph/0702532},
 primaryClass = {astro-ph},
       adsurl = {https://ui.adsabs.harvard.edu/abs/2007A&A...467..611F}
}

@ARTICLE{Fagundes_2012,
       author = {{Fagundes}, Daniel Almeida and {Menon}, Marcio Jos{\'e} and {Silva}, Paulo Victor Recchia Gomes},
        title = "{Total Hadronic Cross-Section Data and the Froissart-Martin Bound}",
      journal = {Brazilian Journal of Physics},
         year = 2012,
        month = dec,
       volume = {42},
       number = {5-6},
        pages = {452-464},
          doi = {10.1007/s13538-012-0099-5},
archivePrefix = {arXiv},
       eprint = {1112.4704},
 primaryClass = {hep-ph},
       adsurl = {https://ui.adsabs.harvard.edu/abs/2012BrJPh..42..452F}
}

@ARTICLE{Weiner_1996,
       author = {{Weiner}, Benjamin J. and {Williams}, T.~B.},
        title = "{Detection of H(alpha) Emission From The Magellanic Stream: Evidence For an Extended Gaseous Galactic Halo}",
      journal = {aj},
         year = 1996,
        month = mar,
       volume = {111},
        pages = {1156},
          doi = {10.1086/117860},
archivePrefix = {arXiv},
       eprint = {astro-ph/9512017},
 primaryClass = {astro-ph},
       adsurl = {https://ui.adsabs.harvard.edu/abs/1996AJ....111.1156W}
}

@article{Kalashev_2023,
doi = {10.1088/1475-7516/2023/03/053},
url = {https://doi.org/10.1088/1475-7516/2023/03/053},
year = {2023},
month = {mar},
publisher = {IOP Publishing},
volume = {2023},
number = {03},
pages = {053},
author = {Kalashev, Oleg and Martynenko, Nickolay and Troitsky, Sergey},
title = {On the contribution of cosmic-ray interactions in the circumgalactic gas to the observed high-energy neutrino flux},
journal = {jcap}
}

@ARTICLE{Mora_2019,
       author = {{Mora-Partiarroyo}, Silvia Carolina and {Krause}, Marita and {Basu}, Aritra and {Beck}, Rainer and {Wiegert}, Theresa and {Irwin}, Judith and {Henriksen}, Richard and {Stein}, Yelena and {Vargas}, Carlos J. and {Heesen}, Volker and {Walterbos}, Ren{\'e} A.~M. and {Rand}, Richard J. and {Heald}, George and {Li}, Jiangtao and {Kamieneski}, Patrick and {English}, Jayanne},
        title = "{CHANG-ES. XIV. Cosmic-ray propagation and magnetic field strengths in the radio halo of NGC 4631}",
      journal = {aap},
         year = 2019,
        month = dec,
       volume = {632},
          eid = {A10},
        pages = {A10},
          doi = {10.1051/0004-6361/201834571},
archivePrefix = {arXiv},
       eprint = {1910.07588},
 primaryClass = {astro-ph.GA},
       adsurl = {https://ui.adsabs.harvard.edu/abs/2019A&A...632A..10M}
}

@ARTICLE{Heesen_2023,
       author = {{Heesen}, V. and {O'Sullivan}, S.~P. and {Br{\"u}ggen}, M. and {Basu}, A. and {Beck}, R. and {Seta}, A. and {Carretti}, E. and {Krause}, M.~G.~H. and {Haverkorn}, M. and {Hutschenreuter}, S. and {Bracco}, A. and {Stein}, M. and {Bomans}, D.~J. and {Dettmar}, R.-J. and {Chy{\.z}y}, K.~T. and {Heald}, G.~H. and {Paladino}, R. and {Horellou}, C.},
        title = "{Detection of magnetic fields in the circumgalactic medium of nearby galaxies using Faraday rotation}",
      journal = {aap},
         year = 2023,
        month = feb,
       volume = {670},
          eid = {L23},
        pages = {L23},
          doi = {10.1051/0004-6361/202346008},
archivePrefix = {arXiv},
       eprint = {2302.06617},
 primaryClass = {astro-ph.GA},
       adsurl = {https://ui.adsabs.harvard.edu/abs/2023A&A...670L..23H}
}

@ARTICLE{Ramesh_2023,
       author = {{Ramesh}, Rahul and {Nelson}, Dylan and {Heesen}, Volker and {Br{\"u}ggen}, Marcus},
        title = "{Azimuthal anisotropy of magnetic fields in the circumgalactic medium driven by galactic feedback processes}",
      journal = {mnras},
         year = 2023,
        month = dec,
       volume = {526},
       number = {4},
        pages = {5483-5493},
          doi = {10.1093/mnras/stad3104},
archivePrefix = {arXiv},
       eprint = {2305.11214},
 primaryClass = {astro-ph.GA},
       adsurl = {https://ui.adsabs.harvard.edu/abs/2023MNRAS.526.5483R}
}

@article{Genolini_2015,
       author = {{Genolini}, Y. and {Putze}, A. and {Salati}, P. and {Serpico}, P.~D.},
        title = "{Theoretical uncertainties in extracting cosmic-ray diffusion parameters: the boron-to-carbon ratio}",
      journal = {aap},
         year = 2015,
        month = aug,
       volume = {580},
          eid = {A9},
        pages = {A9},
          doi = {10.1051/0004-6361/201526344},
archivePrefix = {arXiv},
       eprint = {1504.03134},
 primaryClass = {astro-ph.HE},
       adsurl = {https://ui.adsabs.harvard.edu/abs/2015A&A...580A...9G}
}

@article{dragon_II_2018,
doi = {10.1088/1475-7516/2018/07/006},
url = {https://doi.org/10.1088/1475-7516/2018/07/006},
year = {2018},
month = {jul},
publisher = {},
volume = {2018},
number = {07},
pages = {006},
author = {Evoli, Carmelo and Gaggero, Daniele and Vittino, Andrea and Di Mauro, Mattia and Grasso, Dario and Mazziotta, Mario Nicola},
title = {Cosmic-ray propagation with DRAGON2: II. Nuclear interactions with the interstellar gas},
journal = {jcap},
}

@ARTICLE{Mewaldt_1981,
       author = {{Mewaldt}, R.~A. and {Spalding}, J.~D. and {Stone}, E.~C. and {Vogt}, R.~E.},
        title = "{The isotropic composition of cosmic ray B, C, N, and O nuclei}",
      journal = {apjl},
         year = 1981,
        month = dec,
       volume = {251},
        pages = {L27-L31},
          doi = {10.1086/183686},
       adsurl = {https://ui.adsabs.harvard.edu/abs/1981ApJ...251L..27M}
}

@article{Borchiellini_2026,
title = {Revisiting electron-capture decay for Galactic cosmic-ray data},
journal = {Astroparticle Physics},
volume = {176},
pages = {103203},
year = {2026},
issn = {0927-6505},
doi = {https://doi.org/10.1016/j.astropartphys.2025.103203},
url = {https://www.sciencedirect.com/science/article/pii/S0927650525001264},
author = {M. Borchiellini and D. Maurin and M. Vecchi}
}

@ARTICLE{Yiou_1970,
       author = {{Yiou}, F. and {Raisbeck}, G.~M.},
        title = "{The Decay of $^{7}$Be in Cosmic Rays}",
      journal = {aplett},
         year = 1970,
        month = nov,
       volume = {7},
        pages = {129},
       adsurl = {https://ui.adsabs.harvard.edu/abs/1970ApL.....7..129Y}
}

@ARTICLE{Putze_2010,
       author = {{Putze}, A. and {Derome}, L. and {Maurin}, D.},
        title = "{A Markov Chain Monte Carlo technique to sample transport and source parameters of Galactic cosmic rays. II. Results for the diffusion model combining B/C and radioactive nuclei}",
      journal = {aap},
         year = 2010,
        month = jun,
       volume = {516},
          eid = {A66},
        pages = {A66},
          doi = {10.1051/0004-6361/201014010},
archivePrefix = {arXiv},
       eprint = {1001.0551},
 primaryClass = {astro-ph.HE},
       adsurl = {https://ui.adsabs.harvard.edu/abs/2010A&A...516A..66P}
}

@ARTICLE{Lerche_1982,
       author = {{Lerche}, I. and {Schlickeiser}, R.},
        title = "{On the transport and propagation of cosmic rays in galaxies. I - Solution of the steady-state transport equation for cosmic ray nucleons, momentum spectra and heating of the interstellar medium}",
      journal = {mnras},
         year = 1982,
        month = dec,
       volume = {201},
        pages = {1041-1072},
          doi = {10.1093/mnras/201.4.1041},
       adsurl = {https://ui.adsabs.harvard.edu/abs/1982MNRAS.201.1041L}
}

@ARTICLE{McMillan_2017,
       author = {{McMillan}, Paul J.},
        title = "{The mass distribution and gravitational potential of the Milky Way}",
      journal = {mnras},
         year = 2017,
        month = feb,
       volume = {465},
       number = {1},
        pages = {76-94},
          doi = {10.1093/mnras/stw2759},
archivePrefix = {arXiv},
       eprint = {1608.00971},
 primaryClass = {astro-ph.GA},
       adsurl = {https://ui.adsabs.harvard.edu/abs/2017MNRAS.465...76M}
}

@ARTICLE{Lipari_2022,
       author = {{Lipari}, Paolo},
        title = "{Beryllium isotopic composition and Galactic cosmic ray propagation}",
      journal = {arXiv e-prints},
         year = 2022,
        month = apr,
          eid = {arXiv:2204.13085},
        pages = {arXiv:2204.13085},
          doi = {10.48550/arXiv.2204.13085},
archivePrefix = {arXiv},
       eprint = {2204.13085},
 primaryClass = {astro-ph.HE},
       adsurl = {https://ui.adsabs.harvard.edu/abs/2022arXiv220413085L}
}

@ARTICLE{Lipari_2014,
       author = {{Lipari}, Paolo},
        title = "{The lifetime of cosmic rays in the Milky Way}",
      journal = {arXiv e-prints},
         year = 2014,
        month = jul,
          eid = {arXiv:1407.5223},
        pages = {arXiv:1407.5223},
          doi = {10.48550/arXiv.1407.5223},
archivePrefix = {arXiv},
       eprint = {1407.5223},
 primaryClass = {astro-ph.HE},
       adsurl = {https://ui.adsabs.harvard.edu/abs/2014arXiv1407.5223L}
}

@ARTICLE{Blasi_2013,
       author = {{Blasi}, Pasquale},
        title = "{Origin of Galactic Cosmic Rays}",
      journal = {Nuclear Physics B Proceedings Supplements},
         year = 2013,
        month = jun,
       volume = {239},
        pages = {140-147},
          doi = {10.1016/j.nuclphysbps.2013.05.023},
archivePrefix = {arXiv},
       eprint = {1211.4799},
 primaryClass = {astro-ph.HE},
       adsurl = {https://ui.adsabs.harvard.edu/abs/2013NuPhS.239..140B}
}

@ARTICLE{LHAASO_2025,
       author = {{Lhaaso Collaboration} and {Cao}, Zhen and {Aharonian}, Felix and {Bai}, Yun-Xiang and {Bao}, Yi-Wei and {Bastieri}, Denis and {Bi}, Xiao-Jun and {Bi}, Yu-Jiang and {Bian}, Wen-Yi and {Bukevich}, Anatoly V. and {Cai}, Chengmiao and {Cao}, Wen-Yu and {Cao}, Zhe and {Chang}, Jin and {Chang}, Jin-Fan and {Chen}, Aming and {Chen}, En-Sheng and {Chen}, Guohai and {Chen}, Hua-Xi and {Chen}, Liang and {Chen}, Long and {Chen}, Ming-Jun and {Chen}, Ma-Li and {Chen}, Qi-Hui and {Chen}, Shi and {Chen}, Su-Hong and {Chen}, Song-Zhan and {Chen}, Tian-Lu and {Chen}, Xiao-Bin and {Chen}, Xuejian and {Chen}, Yang and {Cheng}, Ning and {Cheng}, Yao-Dong and {Chung Chu}, Ming and {Cui}, Ming-Yang and {Cui}, Shu-Wang and {Cui}, Xiao-Hong and {Cui}, Yi-Dong and {Dai}, Ben-Zhong and {Dai}, Hong-Liang and {Dai}, Zigao and {Luobu}, Danzeng and {Diao}, Yang-Xuan and {Dong}, Xu-Qiang and {Duan}, Kai-Kai and {Fan}, Jun-Hui and {Fan}, Yi-Zhong and {Fang}, Jun and {Fang}, Jian-Hua and {Fang}, Kun and {Feng}, Cun-Feng and {Feng}, Hua and {Feng}, Li and {Feng}, Shaohui and {Feng}, Xiao-Ting and {Feng}, Yi and {Feng}, You-Liang and {Gabici}, Stefano and {Gao}, Bo and {Gao}, Chuan-Dong and {Gao}, Qi and {Gao}, Wei and {Gao}, Wei-Kang and {Ge}, Maomao and {Ge}, Ting-Ting and {Geng}, Lisi and {Giacinti}, Gwenael and {Gong}, Guanghua and {Gou}, Quanbu and {Gu}, Min-Hao and {Guo}, Fu-Lai and {Guo}, Jing and {Guo}, Xiao-Lei and {Guo}, Yi-Qing and {Guo}, Ying-Ying and {Han}, Yi-Ang and {Hannuksela}, Otto A. and {Hasan}, Mariam and {He}, Hui-Hai and {He}, Hao-Ning and {He}, Jia-Yin and {He}, Xinyu and {He}, Yu and {Hern{\'a}ndez-Cadena}, Sergio and {Hou}, Bo-Wen and {Hou}, Chao and {Hou}, Xian and {Hu}, Hong-Bo and {Hu}, Shi-Cong and {Huang}, Chen and {Huang}, Dai-Hui and {Huang}, Jiajun and {Huang}, Tian-Qi and {Huang}, Wen-Jun and {Huang}, Xing-Tao and {Huang}, Xiao-Yuan and {Huang}, Yong and {Huang}, Yi-Yun and {Ji}, Xiao-Lu and {Jia}, Huan-Yu and {Jia}, Kang and {Jiang}, Hou-Bing and {Jiang}, Kun and {Jiang}, Xiao-Wei and {Jiang}, Ze-Jun and {Jin}, Min and {Kaci}, Samy and {Kang}, Ming-Ming and {Karpikov}, Ivan and {Khangulyan}, Dmitry and {Kuleshov}, Denis and {Kurinov}, Kirill and {Li}, Bing-Bing and {Li}, Cheng and {Li}, Cong and {Li}, Dan and {Li}, Fei and {Li}, Haibo and {Li}, Huicai and {Li}, Jian and {Li}, Jie and {Li}, Kai and {Li}, Long and {Li}, Rong-Lan and {Li}, Si-Da and {Li}, Tian-Yang and {Li}, Wen-Lian and {Li}, Xiu-Rong and {Li}, Xin and {Li}, Yuan and {Li}, Yizhuo and {Li}, Zhe and {Li}, Zhuo and {Liang}, En-Wei and {Liang}, Yun-Feng and {Lin}, Su-Jie and {Liu}, Bing and {Liu}, Cheng and {Liu}, Dong and {Liu}, Dang-Bo and {Liu}, Hu and {Liu}, Hai-Dong and {Liu}, Jia and {Liu}, Jia-Li and {Liu}, Ji-Ren and {Liu}, Mao-Yuan and {Liu}, Ruo-Yu and {Liu}, Si-Ming and {Liu}, Wei and {Liu}, X. and {Liu}, Yi and {Liu}, Yu and {Liu}, Yi-Nong and {Lou}, Yu-Qing and {Luo}, Qing and {Luo}, Yu and {Lv}, Hong-Kui and {Ma}, Bo-Qiang and {Ma}, Ling-Ling and {Ma}, Xin-Hua and {Mao}, Ji-Rong and {Min}, Zhen and {Mitthumsiri}, Warit and {Mou}, Guo-Bin and {Mu}, Hui-Jun and {Neronov}, Andrii and {Ng}, Kenny Chun Yu and {Ni}, Ming-Yang and {Nie}, Lin and {Ou}, Le-Jian and {Pattarakijwanich}, Petchara and {Pei}, Zhi-Yuan and {Qi}, Jin-Can and {Qi}, Meng-Yao and {Qin}, Jia-Jun and {Raza}, Ali and {Ren}, Chong-Yang and {Ruffolo}, David and {S{\'a}iz}, Alejandro and {Semikoz}, Dmitri and {Shao}, Lang and {Shchegolev}, Oleg and {Shen}, Yun-Zhi and {Sheng}, Xiang-Dong and {Shi}, Zhaodong and {Shu}, Fu-Wen and {Song}, Hui-Chao and {Stenkin}, Yuri V. and {Stepanov}, Vladimir and {Su}, Yang and {Sun}, Dongxu and {Sun}, Hao and {Sun}, Qinning and {Sun}, Xiaona and {Sun}, Zhibin and {Hussain Tabasam}, Nabeel and {Takata}, Jumpei and {Tam}, Pak Hin Thomas and {Tan}, Hong-Bin and {Tang}, Qingwen},
        title = "{Ultrahigh-Energy Gamma-ray Emission Associated with Black Hole-Jet Systems}",
      journal = {National Science Review},
         year = 2025,
        month = dec,
       volume = {12},
       number = {12},
          eid = {nwaf496},
        pages = {nwaf496},
          doi = {10.1093/nsr/nwaf496},
archivePrefix = {arXiv},
       eprint = {2410.08988},
 primaryClass = {astro-ph.HE},
       adsurl = {https://ui.adsabs.harvard.edu/abs/2025NSRev..12af496L}
}

@article{Mengjie_2024,
  title = {Interpretation of AMS-02 beryllium isotope fluxes using data-driven production cross sections},
  author = {Zhao, Meng-Jie and Bi, Xiao-Jun and Fang, Kun and Yin, Peng-Fei},
  journal = {prd},
  volume = {109},
  issue = {8},
  pages = {083036},
  numpages = {15},
  year = {2024},
  month = {Apr},
  publisher = {American Physical Society},
  doi = {10.1103/PhysRevD.109.083036},
  url = {https://link.aps.org/doi/10.1103/PhysRevD.109.083036}
}

@article{Silver_2024,
doi = {10.3847/1538-4357/ad1ce8},
url = {https://doi.org/10.3847/1538-4357/ad1ce8},
year = {2024},
month = {mar},
publisher = {The American Astronomical Society},
volume = {963},
number = {2},
pages = {111},
author = {Silver, Ethan and Orlando, Elena},
title = {Testing Cosmic-Ray Propagation Scenarios with AMS-02 and Voyager Data},
journal = {apj}
}

@article{Korsmeier_2021,
  title = {Implications of lithium to oxygen AMS-02 spectra on our understanding of cosmic-ray diffusion},
  author = {Korsmeier, Michael and Cuoco, Alessandro},
  journal = {prd},
  volume = {103},
  issue = {10},
  pages = {103016},
  numpages = {24},
  year = {2021},
  month = {May},
  publisher = {American Physical Society},
  doi = {10.1103/PhysRevD.103.103016},
  url = {https://link.aps.org/doi/10.1103/PhysRevD.103.103016}
}

@article{TorreLuque_2021,
doi = {10.1088/1475-7516/2021/03/099},
url = {https://doi.org/10.1088/1475-7516/2021/03/099},
year = {2021},
month = {mar},
publisher = {IOP Publishing},
volume = {2021},
number = {03},
pages = {099},
author = {De La Torre Luque, P. and Mazziotta, M.N. and Loparco, F. and Gargano, F. and Serini, D.},
title = {Implications of current nuclear cross sections on secondary cosmic rays with the upcoming DRAGON2 code},
journal = {jcap}
}

@article{Weinrich_2020,
	author = {{Weinrich, N.} and {G\'enolini, Y.} and {Boudaud, M.} and {Derome, L.} and {Maurin, D.}},
	title = {Combined analysis of AMS-02 (Li,Be,B)/C, N/O, 3He, and 4He data},
	DOI= {10.1051/0004-6361/202037875},
	url= {https://doi.org/10.1051/0004-6361/202037875},
	journal = {A\&A},
	year = {2020},
	volume = {639},
	pages = "A131",
}

@article{LHAASO_2023,
  title = {Measurement of Ultra-High-Energy Diffuse Gamma-Ray Emission of the Galactic Plane from 10 TeV to 1 PeV with LHAASO-KM2A},
  author = {Cao, Zhen and Aharonian, F. and An, Q. and Axikegu and Bai, Y. X. and Bao, Y. W. and Bastieri, D. and Bi, X. J. and Bi, Y. J. and Cai, J. T. and Cao, Q. and Cao, W. Y. and Cao, Zhe and Chang, J. and Chang, J. F. and Chen, A. M. and Chen, E. S. and Chen, Liang and Chen, Lin and Chen, Long and Chen, M. J. and Chen, M. L. and Chen, Q. H. and Chen, S. H. and Chen, S. Z. and Chen, T. L. and Chen, Y. and Cheng, N. and Cheng, Y. D. and Cui, M. Y. and Cui, S. W. and Cui, X. H. and Cui, Y. D. and Dai, B. Z. and Dai, H. L. and Dai, Z. G. and Danzengluobu and della Volpe, D. and Dong, X. Q. and Duan, K. K. and Fan, J. H. and Fan, Y. Z. and Fang, J. and Fang, K. and Feng, C. F. and Feng, L. and Feng, S. H. and Feng, X. T. and Feng, Y. L. and Gabici, S. and Gao, B. and Gao, C. D. and Gao, L. Q. and Gao, Q. and Gao, W. and Gao, W. K. and Ge, M. M. and Geng, L. S. and Giacinti, G. and Gong, G. H. and Gou, Q. B. and Gu, M. H. and Guo, F. L. and Guo, X. L. and Guo, Y. Q. and Guo, Y. Y. and Han, Y. A. and He, H. H. and He, H. N. and He, J. Y. and He, X. B. and He, Y. and Heller, M. and Hor, Y. K. and Hou, B. W. and Hou, C. and Hou, X. and Hu, H. B. and Hu, Q. and Hu, S. C. and Huang, D. H. and Huang, T. Q. and Huang, W. J. and Huang, X. T. and Huang, X. Y. and Huang, Y. and Huang, Z. C. and Ji, X. L. and Jia, H. Y. and Jia, K. and Jiang, K. and Jiang, X. W. and Jiang, Z. J. and Jin, M. and Kang, M. M. and Ke, T. and Kuleshov, D. and Kurinov, K. and Li, B. B. and Li, Cheng and Li, Cong and Li, D. and Li, F. and Li, H. B. and Li, H. C. and Li, H. Y. and Li, J. and Li, Jian and Li, Jie and Li, K. and Li, W. L. and Li, W. L. and Li, X. R. and Li, Xin and Li, Y. Z. and Li, Zhe and Li, Zhuo and Liang, E. W. and Liang, Y. F. and Lin, S. J. and Liu, B. and Liu, C. and Liu, D. and Liu, H. and Liu, H. D. and Liu, J. and Liu, J. L. and Liu, J. Y. and Liu, M. Y. and Liu, R. Y. and Liu, S. M. and Liu, W. and Liu, Y. and Liu, Y. N. and Lu, R. and Luo, Q. and Lv, H. K. and Ma, B. Q. and Ma, L. L. and Ma, X. H. and Mao, J. R. and Min, Z. and Mitthumsiri, W. and Mu, H. J. and Nan, Y. C. and Neronov, A. and Ou, Z. W. and Pang, B. Y. and Pattarakijwanich, P. and Pei, Z. Y. and Qi, M. Y. and Qi, Y. Q. and Qiao, B. Q. and Qin, J. J. and Ruffolo, D. and S\'aiz, A. and Semikoz, D. and Shao, C. Y. and Shao, L. and Shchegolev, O. and Sheng, X. D. and Shu, F. W. and Song, H. C. and Stenkin, Yu. V. and Stepanov, V. and Su, Y. and Sun, Q. N. and Sun, X. N. and Sun, Z. B. and Tam, P. H. T. and Tang, Q. W. and Tang, Z. B. and Tian, W. W. and Wang, C. and Wang, C. B. and Wang, G. W. and Wang, H. G. and Wang, H. H. and Wang, J. C. and Wang, K. and Wang, L. P. and Wang, L. Y. and Wang, P. H. and Wang, R. and Wang, W. and Wang, X. G. and Wang, X. Y. and Wang, Y. and Wang, Y. D. and Wang, Y. J. and Wang, Z. H. and Wang, Z. X. and Wang, Zhen and Wang, Zheng and Wei, D. M. and Wei, J. J. and Wei, Y. J. and Wen, T. and Wu, C. Y. and Wu, H. R. and Wu, S. and Wu, X. F. and Wu, Y. S. and Xi, S. Q. and Xia, J. and Xia, J. J. and Xiang, G. M. and Xiao, D. X. and Xiao, G. and Xin, G. G. and Xin, Y. L. and Xing, Y. and Xiong, Z. and Xu, D. L. and Xu, R. F. and Xu, R. X. and Xu, W. L. and Xue, L. and Yan, D. H. and Yan, J. Z. and Yan, T. and Yang, C. W. and Yang, F. and Yang, F. F. and Yang, H. W. and Yang, J. Y. and Yang, L. L. and Yang, M. J. and Yang, R. Z. and Yang, S. B. and Yao, Y. H. and Yao, Z. G. and Ye, Y. M. and Yin, L. Q. and Yin, N. and You, X. H. and You, Z. Y. and Yu, Y. H. and Yuan, Q. and Yue, H. and Zeng, H. D. and Zeng, T. X. and Zeng, W. and Zha, M. and Zhang, B. B. and Zhang, F. and Zhang, H. M. and Zhang, H. Y. and Zhang, J. L. and Zhang, L. X. and Zhang, Li and Zhang, P. F. and Zhang, P. P. and Zhang, R. and Zhang, S. B. and Zhang, S. R. and Zhang, S. S. and Zhang, X. and Zhang, X. P. and Zhang, Y. F. and Zhang, Yi and Zhang, Yong and Zhao, B. and Zhao, J. and Zhao, L. and Zhao, L. Z. and Zhao, S. P. and Zheng, F. and Zhou, B. and Zhou, H. and Zhou, J. N. and Zhou, M. and Zhou, P. and Zhou, R. and Zhou, X. X. and Zhu, C. G. and Zhu, F. R. and Zhu, H. and Zhu, K. J. and Zuo, X.},
  collaboration = {LHAASO Collaboration},
  journal = {prl},
  volume = {131},
  issue = {15},
  pages = {151001},
  numpages = {9},
  year = {2023},
  month = {Oct},
  publisher = {American Physical Society},
  doi = {10.1103/PhysRevLett.131.151001},
  url = {https://link.aps.org/doi/10.1103/PhysRevLett.131.151001}
}

@article{IceCube2023,
author = {{IceCube Collaboration} and R. Abbasi  and M. Ackermann  and J. Adams  and J. A. Aguilar  and M. Ahlers  and M. Ahrens  and J. M. Alameddine  and A. A. Alves  and N. M. Amin  and K. Andeen  and T. Anderson  and G. Anton  and C. Argüelles  and Y. Ashida  and S. Athanasiadou  and S. Axani  and X. Bai  and A. Balagopal V.  and S. W. Barwick  and V. Basu  and S. Baur  and R. Bay  and J. J. Beatty  and K.-H. Becker  and J. Becker Tjus  and J. Beise  and C. Bellenghi  and S. Benda  and S. BenZvi  and D. Berley  and E. Bernardini  and D. Z. Besson  and G. Binder  and D. Bindig  and E. Blaufuss  and S. Blot  and M. Boddenberg  and F. Bontempo  and J. Y. Book  and J. Borowka  and S. Böser  and O. Botner  and J. Böttcher  and E. Bourbeau  and F. Bradascio  and J. Braun  and B. Brinson  and S. Bron  and J. Brostean-Kaiser  and R. T. Burley  and R. S. Busse  and M. A. Campana  and E. G. Carnie-Bronca  and C. Chen  and Z. Chen  and D. Chirkin  and K. Choi  and B. A. Clark  and K. Clark  and L. Classen  and A. Coleman  and G. H. Collin  and A. Connolly  and J. M. Conrad  and P. Coppin  and P. Correa  and D. F. Cowen  and R. Cross  and C. Dappen  and P. Dave  and C. De Clercq  and J. J. DeLaunay  and D. Delgado López  and H. Dembinski  and K. Deoskar  and A. Desai  and P. Desiati  and K. D. de Vries  and G. de Wasseige  and T. DeYoung  and A. Diaz  and J. C. Díaz-Vélez  and M. Dittmer  and H. Dujmovic  and M. Dunkman  and M. A. DuVernois  and T. Ehrhardt  and P. Eller  and R. Engel  and H. Erpenbeck  and J. Evans  and P. A. Evenson  and K. L. Fan  and A. R. Fazely  and A. Fedynitch  and N. Feigl  and S. Fiedlschuster  and A. T. Fienberg  and C. Finley  and L. Fischer  and D. Fox  and A. Franckowiak  and E. Friedman  and A. Fritz  and P. Fürst  and T. K. Gaisser  and J. Gallagher  and E. Ganster  and A. Garcia  and S. Garrappa  and L. Gerhardt  and A. Ghadimi  and C. Glaser  and T. Glauch  and T. Glüsenkamp  and N. Goehlke  and A. Goldschmidt  and J. G. Gonzalez  and S. Goswami  and D. Grant  and T. Grégoire  and S. Griswold  and C. Günther  and P. Gutjahr  and C. Haack  and A. Hallgren  and R. Halliday  and L. Halve  and F. Halzen  and M. Ha Minh  and K. Hanson  and J. Hardin  and A. A. Harnisch  and A. Haungs  and K. Helbing  and F. Henningsen  and E. C. Hettinger  and S. Hickford  and J. Hignight  and C. Hill  and G. C. Hill  and K. D. Hoffman  and K. Hoshina  and W. Hou  and F. Huang  and M. Huber  and T. Huber  and K. Hultqvist  and M. Hünnefeld  and R. Hussain  and K. Hymon  and S. In  and N. Iovine  and A. Ishihara  and M. Jansson  and G. S. Japaridze  and M. Jeong  and M. Jin  and B. J. P. Jones  and D. Kang  and W. Kang  and X. Kang  and A. Kappes  and D. Kappesser  and L. Kardum  and T. Karg  and M. Karl  and A. Karle  and U. Katz  and M. Kauer  and M. Kellermann  and J. L. Kelley  and A. Kheirandish  and K. Kin  and J. Kiryluk  and S. R. Klein  and A. Kochocki  and R. Koirala  and H. Kolanoski  and T. Kontrimas  and L. Köpke  and C. Kopper  and S. Kopper  and D. J. Koskinen  and P. Koundal  and M. Kovacevich  and M. Kowalski  and T. Kozynets  and E. Krupczak  and E. Kun  and N. Kurahashi  and N. Lad  and C. Lagunas Gualda  and J. L. Lanfranchi  and M. J. Larson  and F. Lauber  and J. P. Lazar  and J. W. Lee  and K. Leonard  and A. Leszczyńska  and Y. Li  and M. Lincetto  and Q. R. Liu  and M. Liubarska  and E. Lohfink  and C. J. Lozano Mariscal  and L. Lu  and F. Lucarelli  and A. Ludwig  and W. Luszczak  and Y. Lyu  and W. Y. Ma  and J. Madsen  and K. B. M. Mahn  and Y. Makino  and S. Mancina  and I. C. Mariş  and I. Martinez-Soler  and R. Maruyama  and S. McHale  and T. McElroy  and F. McNally  and J. V. Mead  and K. Meagher  and S. Mechbal  and A. Medina  and M. Meier  and S. Meighen-Berger  and Y. Merckx  and J. Micallef  and D. Mockler  and T. Montaruli  and R. W. Moore  and K. Morik  and R. Morse  and M. Moulai  and T. Mukherjee  and R. Naab  and R. Nagai  and R. Nahnhauer  and U. Naumann  and J. Necker  and L. V. Nguyen  and H. Niederhausen  and M. U. Nisa  and S. C. Nowicki  and D. Nygren  and A. Obertacke Pollmann  and M. Oehler  and B. Oeyen  and A. Olivas  and E. O'Sullivan  and H. Pandya  and D. V. Pankova  and N. Park  and G. K. Parker  and E. N. Paudel  and L. Paul  and C. Pérez de los Heros  and L. Peters  and J. Peterson  and S. Philippen  and S. Pieper  and A. Pizzuto  and M. Plum  and Y. Popovych  and A. Porcelli  and M. Prado Rodriguez  and B. Pries  and G. T. Przybylski  and C. Raab  and J. Rack-Helleis  and A. Raissi  and M. Rameez  and K. Rawlins  and I. C. Rea  and Z. Rechav  and A. Rehman  and P. Reichherzer  and R. Reimann  and G. Renzi  and E. Resconi  and S. Reusch  and W. Rhode  and M. Richman  and B. Riedel  and E. J. Roberts  and S. Robertson  and G. Roellinghoff  and M. Rongen  and C. Rott  and T. Ruhe  and D. Ryckbosch  and D. Rysewyk Cantu  and I. Safa  and J. Saffer  and D. Salazar-Gallegos  and P. Sampathkumar  and S. E. Sanchez Herrera  and A. Sandrock  and M. Santander  and S. Sarkar  and S. Sarkar  and K. Satalecka  and M. Schaufel  and H. Schieler  and S. Schindler  and T. Schmidt  and A. Schneider  and J. Schneider  and F. G. Schröder  and L. Schumacher  and G. Schwefer  and S. Sclafani  and D. Seckel  and S. Seunarine  and A. Sharma  and S. Shefali  and N. Shimizu  and M. Silva  and B. Skrzypek  and B. Smithers  and R. Snihur  and J. Soedingrekso  and A. Sogaard  and D. Soldin  and C. Spannfellner  and G. M. Spiczak  and C. Spiering  and M. Stamatikos  and T. Stanev  and R. Stein  and J. Stettner  and T. Stezelberger  and B. Stokstad  and T. Stürwald  and T. Stuttard  and G. W. Sullivan  and I. Taboada  and S. Ter-Antonyan  and J. Thwaites  and S. Tilav  and F. Tischbein  and K. Tollefson  and C. Tönnis  and S. Toscano  and D. Tosi  and A. Trettin  and M. Tselengidou  and C. F. Tung  and A. Turcati  and R. Turcotte  and C. F. Turley  and J. P. Twagirayezu  and B. Ty  and M. A. Unland Elorrieta  and N. Valtonen-Mattila  and J. Vandenbroucke  and N. van Eijndhoven  and D. Vannerom  and J. van Santen  and J. Veitch-Michaelis  and S. Verpoest  and C. Walck  and W. Wang  and T. B. Watson  and C. Weaver  and P. Weigel  and A. Weindl  and M. J. Weiss  and J. Weldert  and C. Wendt  and J. Werthebach  and M. Weyrauch  and N. Whitehorn  and C. H. Wiebusch  and N. Willey  and D. R. Williams  and M. Wolf  and G. Wrede  and J. Wulff  and X. W. Xu  and J. P. Yanez  and E. Yildizci  and S. Yoshida  and S. Yu  and T. Yuan  and Z. Zhang  and P. Zhelnin },
collaboration = {IceCube Collaboration},
title = {Observation of high-energy neutrinos from the Galactic plane},
journal = {Science},
volume = {380},
number = {6652},
pages = {1338-1343},
year = {2023},
doi = {10.1126/science.adc9818},
URL = {https://www.science.org/doi/abs/10.1126/science.adc9818},
eprint = {https://www.science.org/doi/pdf/10.1126/science.adc9818}
}

@ARTICLE{Zhang_2026,
       author = {{Zhang}, Yi and {Shreeram}, Soumya and {Ponti}, Gabriele and {Comparat}, Johan and {Merloni}, Andrea and {Qu}, Zhijie and {Li}, Jiangtao and {Bregman}, Joel N. and {Fang}, Taotao},
        title = "{On the baryon budget in the X-ray-emitting circumgalactic medium of Milky Way-mass galaxies}",
      journal = {aap},
         year = 2026,
        month = feb,
       volume = {706},
          eid = {A102},
        pages = {A102},
          doi = {10.1051/0004-6361/202556835},
archivePrefix = {arXiv},
       eprint = {2511.17313},
 primaryClass = {astro-ph.GA},
       adsurl = {https://ui.adsabs.harvard.edu/abs/2026A&A...706A.102Z}
}

@ARTICLE{Pshirkov_2026,
       author = {{Pshirkov}, M.~S. and {Nizamov}, B.~A.},
        title = "{Detection of Gamma-Ray Halos around Nearby Late-Type Galaxies}",
      journal = {prl},
         year = 2026,
        month = feb,
       volume = {136},
       number = {8},
          eid = {081201},
        pages = {081201},
          doi = {10.1103/kfld-35xl},
archivePrefix = {arXiv},
       eprint = {2410.02066},
 primaryClass = {astro-ph.HE},
       adsurl = {https://ui.adsabs.harvard.edu/abs/2026PhRvL.136h1201P}
}

@ARTICLE{Recchia_2021,
       author = {{Recchia}, S. and {Gabici}, S. and {Aharonian}, F.~A. and {Niro}, V.},
        title = "{Giant Cosmic-Ray Halos around M31 and the Milky Way}",
      journal = {apj},
         year = 2021,
        month = jun,
       volume = {914},
       number = {2},
          eid = {135},
        pages = {135},
          doi = {10.3847/1538-4357/abfda4},
archivePrefix = {arXiv},
       eprint = {2101.05016},
 primaryClass = {astro-ph.HE},
       adsurl = {https://ui.adsabs.harvard.edu/abs/2021ApJ...914..135R}
}

@ARTICLE{Morrison_1954,
       author = {{Morrison}, Philip and {Olbert}, Stanislaw and {Rossi}, Bruno},
        title = "{The Origin of Cosmic Rays}",
      journal = {Physical Review},
         year = 1954,
        month = apr,
       volume = {94},
       number = {2},
        pages = {440-453},
          doi = {10.1103/PhysRev.94.440},
       adsurl = {https://ui.adsabs.harvard.edu/abs/1954PhRv...94..440M}
}
\bibliographystyle{aasjournal}



\end{document}